\documentclass[preprint,12pt]{elsarticle}

\usepackage{graphicx}
\usepackage{amsmath}

\usepackage{amssymb}
\biboptions{sort&compress}

\usepackage[utf8]{inputenc}

\usepackage[htt]{hyphenat}

\usepackage{listings}
\usepackage{hyperref}
\usepackage{xcolor}%,colordvi}%,pifont,times
\definecolor{links}{HTML}{0000EE}%{2A1B81}
\hypersetup{colorlinks,linkcolor=,urlcolor=links}

\newcommand{\bvar}{\mathbf}

\usepackage{xspace}
\newcommand{\thermalfist}{\textsc{Thermal-FIST}\xspace}

\newcommand{\eq}[1]{\begin{align} #1 \end{align}}

\newcounter{bla}

\journal{Computer Physics Communications}

\begin{document}

\begin{frontmatter}



\title{
\begin{center}
\includegraphics[width=0.15\textwidth]{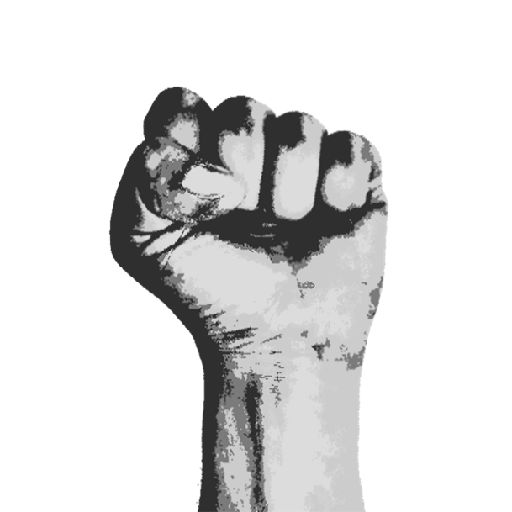}
\end{center}
\thermalfist: A package for heavy-ion collisions and hadronic equation of state}

\author[a,b]{Volodymyr Vovchenko\corref{author}}
\author[a,b,c]{Horst Stoecker}

\cortext[author] {Corresponding author.}%\\\textit{E-mail address:}}
\address[a]{Institut f\"ur Theoretische Physik,
Goethe Universit\"at Frankfurt,\\ D-60438 Frankfurt am Main, Germany}
\address[b]{Frankfurt Institute for Advanced Studies, Goethe Universit\"at Frankfurt,\\
D-60438 Frankfurt am Main, Germany}
\address[c]{GSI Helmholtzzentrum f\"ur Schwerionenforschung GmbH,\\ D-64291 Darmstadt, Germany}

\renewcommand*{\thefootnote}{\fnsymbol{footnote}}

\begin{abstract}
\thermalfist\footnote[7]{Thermal-FIST -- Thermal, Fast and Interactive Statistical Toolkit} is a C++ package designed for convenient general-purpose physics analysis within the family of hadron resonance gas (HRG) models.
This mainly includes the statistical analysis of particle production in heavy-ion collisions and the phenomenology of  hadronic equation of state. 
Notable features include fluctuations and correlations of conserved charges, effects of probabilistic decay, chemical non-equilibrium, and inclusion of van der Waals hadronic interactions.
Calculations are possible within the grand canonical ensemble, the canonical ensemble, as well as in mixed-canonical ensembles combining the canonical treatment of certain conserved charges with the grand-canonical treatment of other conserved charges.
The package contains a fast thermal event generator, which generates particle yields in accordance with the HRG chemistry, and particle momenta based on the Blast Wave model.
A distinct feature of this package is the presence of the graphical user interface frontend -- \textsc{QtThermalFIST} -- which is designed for fast and convenient general-purpose HRG model applications.

\end{abstract}

\renewcommand*{\thefootnote}{\arabic{footnote}}
\setcounter{footnote}{0}


\end{frontmatter}

% \linenumbers

% Computer program descriptions should contain the following
% PROGRAM SUMMARY.

{\bf Program summary}
  %Delete as appropriate.

\begin{small}
\noindent
{\em Program Title:} \thermalfist, version 1.2                                          \\
{\em Licensing provisions:} GPLv30                                    \\
{\em Programming language:} C++                                   \\
{\em Computer:} any with a C++ compiler and, optionally, the Qt5 framework for the GUI frontend                                   \\
{\em Operating system:} cross-platform, tested on Linux Ubuntu 16.04, 18.04; Mac OS X Yosemite,  Microsoft Windows 10; Android 8, 9                                 \\
{\em External routines:} \href{http://eigen.tuxfamily.org}{Eigen} template library for the linear algebra routines~\cite{Eigen}, \href{https://project-mathlibs.web.cern.ch/project-mathlibs/sw/Minuit2/html/}{MINUIT2} package from CERN ROOT~\cite{MINUIT2}, 
Mersenne Twister random number generator~\cite{MersenneTwister},
\href{https://www.qt.io}{Qt5 framework}~\cite{Qt} (for the GUI only), 
\href{https://www.qcustomplot.com/}{QCustomPlot} Qt widget~\cite{QCustomPlot} (for the GUI only)\\
%{\em Supplementary material:}                                 \\
  % Fill in if necessary, otherwise leave out.
{\em Nature of problem:} 
\\
  The HRG model and its various modifications constitute a common framework used for modeling of the hadronic equation of state and particle production in heavy-ion collisions. 
Even the simplest versions of the HRG model require careful considerations of the many details, including the resonance decay feed-down, implementation of charge conservation constraints relevant for heavy-ion collisions, chemical non-equilibrium effects. 
A notable extra effort is required in order to treat the fluctuations and correlations of various charges charges, which presently are  being extensively studied in the heavy-ion collision experiments and lattice QCD calculations.
The inclusion of hadronic interactions, modeled by an excluded-volume (EV) or a van der Waals~(vdW) type framework, additionally requires a numerical solution to a system of many transcendental equations. 
\\
{\em Solution method:}
\\
The \thermalfist package contains a class-based library which calculates relevant HRG observables for a specified setup. The setup includes a particle list, usually to be supplied with an external file, an HRG model specification (statistical ensemble, van der Waals interaction parameters, etc.), a set of thermal parameters, and conservation laws constraints. Whenever necessary, the systems of transcendental equations are solved numerically with the Broyden's method.
The package includes a fitter for extracting thermal parameters from  hadron yield data through the $\chi^2$ minimization. 
The HRG model based Monte Carlo event generator is a complementary feature to analytic calculations.
A general-purpose thermal analysis is made maximally convenient with QtThermalFIST -- a GUI frontend based on the Qt framework where all typical calculations, such as the properties of the equation of state or the thermal fits, can be straightforwardly performed.
\\
{\em Additional comments:
%including Restrictions and Unusual features:
}\\
If the EV/vdW interactions are present, exact analytic calculations are presently only possible within the grand canonical ensemble.
Approximate calculations are possible for the strangeness-canonical ensemble on the condition that strange particles form a small subsystem relative to the total system.
Effects of probabilistic decays on fluctuation observables are generally included only up to the moments of the 2nd order.
The only exception is the ideal HRG model in the grand canonical ensemble, where these effects are included up to the moments of the 4th order.
On the other hand, the Monte Carlo event generator is not constrained by the above restrictions. 
\\
{\em Running time:}\\
Depending on the specific task, calculation times may vary from milliseconds (an evaluation of thermodynamic functions within the ideal HRG model in the GCE) to several minutes (a thermal fit with sophisticated features such as energy-dependent widths, van der Waals interactions, or exact charge conservation).
The thermal event generator takes few milliseconds to generate an event with about a thousand particles in the GCE.
Event generation in the CE is considerably slower and depends on the rejection sampling rate realized for the given setup.

\begingroup
\renewcommand{\section}[2]{}%

\endgroup

% * Items marked with an asterisk are only required for new versions
% of programs previously published in the CPC Program Library.\\
\end{small}

\section{Introduction}
\label{sec:intro}

The abundant hadron production in heavy-ion collision reactions has long been treated in the framework of the thermal-statistical models~\cite{Mekjian:1977ei,Gosset:1988na,Mekjian:1978us,Stoecker:1981za,Csernai:1986qf,Hahn:1987tz,Hahn:1986mb,Cleymans:1992zc,BraunMunzinger:1996mq,Becattini:2000jw}.
Such a description assumes emission of particles from a thermally and chemically equilibrated source created in these reactions.
Fitting the observed yields of stable hadrons allows to determine the thermal parameters, corresponding to the so-called chemical freeze-out stage of the collision.
In most cases, the ideal hadron resonance gas (Id-HRG) model has been used,
and a surprisingly good description of many experimental hadron yield data
from heavy-ion collisions have been achieved within this
simple approach for a broad range of collision energies, ranging from moderate energies at the SchwerIonen-Synchrotron (SIS) to the highest energy of
the Large Hadron Collider
(LHC)~(see, e.g., Refs.~\cite{Letessier:2005qe,Becattini:2009sc,Andronic:2017pug} for an overview).

HRG-type models also play an important role in the phenomenology of the QCD equation of state. The Id-HRG model is a popular choice for describing the low temperature, confined phase of QCD.
At temperatures between $T \sim 100-150$~MeV and at zero chemical potential, the Id-HRG model indeed
appears to reproduce many lattice QCD observables~\cite{Borsanyi:2011sw,Bazavov:2012jq,Bellwied:2015lba,Bellwied:2013cta}.

Several implementations of the Id-HRG model exist on the market,
including \textsc{SHARE}~\cite{Torrieri:2004zz,Torrieri:2006xi,Petran:2013dva},
\textsc{THERMUS}~\cite{Wheaton:2004qb}, and \textsc{THERMINATOR}~\cite{Kisiel:2005hn,Chojnacki:2011hb}.

In many applications, significant deviations from the ideal gas picture can be expected. 
Extensions of the ideal gas picture have been
discussed mostly within the excluded volume (EV) HRG model~\cite{Rischke:1991ke,Yen:1997rv,Yen:1998pa}, where the
effects of repulsive hadronic interactions at
short distances are introduced (see, e.g., \cite{Satarov:2016peb} for recent developments).
Another extension is the quantum van der Waals (QvdW) model~\cite{Vovchenko:2015vxa,Vovchenko:2016rkn,Vovchenko:2017zpj}, which allows to include both the repulsive and attractive interactions between hadrons.
Recently, repulsive interactions have received renewed interest in the context of lattice QCD data on fluctuations of conserved charges.
It was indicated that large deviations of several fluctuation observables from the Id-HRG baseline are captured by HRG models with repulsive baryon-baryon interactions~\cite{Vovchenko:2016rkn,Huovinen:2017ogf,Vovchenko:2017xad,Vovchenko:2017drx}.
The \thermalfist package presented here allows to calculate thermodynamic features and fluctuation observables within a HRG with arbitrary attractive and repulsive QvdW parameters characterizing interactions between each type of hadron species.

Possible effects of chemical non-equilibrium for light, strange, and/or charm quarks in heavy-ion collisions have been studied in the framework of the Id-HRG model within the SHARE package~\cite{Letessier:2005qe,Petran:2013lja}.
The \thermalfist package allows to study the chemical non-equilibrium effects simultaneously with the effects of EV/vdW interactions or exact charge conservation within the canonical ensemble.

Recent thermal model applications also include the description of the measurements of multiplicity fluctuations~\cite{Garg:2013ata,Fu:2013gga,Alba:2014eba,Alba:2015iva}.
Treatment of probabilistic decays~\cite{Nahrgang:2014fza} and hadronic interactions~\cite{Fu:2013gga} is important in such applications, and it is included into the \thermalfist package.

The applications of the \thermalfist package which can be found in the literature include chemical freeze-out analysis in proton-proton~\cite{Vovchenko:2015idt,Begun:2018qkw} and nucleus-nucleus collisions~\cite{Begun:2018efg,Vovchenko:2018fmh,Motornenko:2018gdc},
influence of the EV interactions on thermal fits~\cite{Vovchenko:2015cbk,Vovchenko:2016ebv,Alba:2016hwx,Satarov:2016peb}
and equation of state~\cite{Anchishkin:2014hfa,Vovchenko:2014pka,Vovchenko:2017xad,Vovchenko:2017drx}, Monte Carlo analysis of EV effects in the canonical ensemble~\cite{Vovchenko:2018cnf},
effects of the vdW interactions between baryons and nuclear liquid-gas transition on various observables~\cite{Vovchenko:2016rkn,Vovchenko:2017cbu,Vovchenko:2017zpj,Vovchenko:2017ygz,Vovchenko:2017ayq}, and thermal production of light nuclei~\cite{Vovchenko:2016mwg,Vovchenko:2018fiy}.

\section{Hadron resonance gas}
\label{sec:HRG}

\subsection{Ideal HRG}

In the simplest setup, the thermodynamics of hadronic phase is described by a multi-component,
ideal gas of point-like hadrons -- the Id-HRG model.
In the grand canonical ensemble (GCE) formulation of the Id-HRG model there are no correlations between different hadronic species. Thus, the pressure is given by
\eq{\label{eq:PHRGid}
p(T,\mu) 
= \sum_i p_i^{\rm id}(T,\mu_i),
}
where the sum goes over all hadron species included in the model,
$p^{\rm id}_i (T, \mu_i)$ is the pressure of the ideal 
Fermi or Bose 
gas at the corresponding temperature and chemical potential for species $i$:
\eq{\label{eq:piid}
p^{\rm id}_i (T, \mu_i) = \frac{d_i}{6\pi^2} \int_0^{\infty} \frac{k^4 dk}{\sqrt{k^2+m_i^2}}
\left[ \exp\left(\frac{\sqrt{k^2+m_i^2} - \mu_i}{T}\right)+\eta_i\right]^{-1}~,
}
where $d_i$ and $m_i$ are, respectively, the spin degeneracy factor and mass of hadron species $i$, and where $\eta_i$ equals +1 for fermions, -1 for bosons, and 0 for the Boltzmann approximation.

Other thermodynamic quantities are given by expressions similar to~\eqref{eq:PHRGid}: a sum over the corresponding ideal gas quantities for all hadron species. 
The particle density of hadron species $i$ is $n_i^{\rm id}(T,\mu_i)$, i.e. it is simply given by
the ideal gas relation for species $i$:
\eq{\label{eq:niid}
n^{\rm id}_i (T, \mu_i) = 
%\left(
\frac{\partial p^{\rm id}_i (T, \mu_i)}{\partial \mu_i}
%\right)_T 
=  \frac{d_i}{2\pi^2} \int_0^{\infty} k^2 dk
\left[ \exp\left(\frac{\sqrt{k^2+m_i^2} - \mu_i}{T}\right)+\eta_i\right]^{-1}~,
}
while the energy density is $\varepsilon(T,\mu) = \sum_i \varepsilon_i^{\rm id}(T,\mu_i)$ with
\eq{\label{eq:eiid}
\varepsilon^{\rm id}_i (T, \mu_i) = 
 \frac{d_i}{2\pi^2} \int_0^{\infty} k^2 \, dk \, \sqrt{k^2 + m_i^2} 
\left[ \exp\left(\frac{\sqrt{k^2+m_i^2} - \mu_i}{T}\right)+\eta_i\right]^{-1}~.
}

Within the GCE formulation, all conserved charges, such as
baryonic number $B$, electric charge $Q$, strangeness $S$, and charm $C$,
are conserved on average.
For the chemical equilibrium case these average values are regulated by the corresponding independent chemical potentials: $\mu_B$, $\mu_Q$, $\mu_S$, and $\mu_C$\footnote{Other charges, such a bottomness, can be considered additionally, in the same manner.}. The chemical potential of the $i$th hadron species is thus determined as 
\begin{equation}
\mu_i\ =\ B_i\,\mu_B\, +\, S_i\,\mu_S\, +\, Q_i\,\mu_Q\, + \, C_i \, \mu_C
\label{eq:mui}
\end{equation}
with $B_i = 0,\, \pm 1$, $S_i = 0,\, \pm 1,\, \pm 2,\, \pm 3$,
$Q_i = 0,\, \pm 1,\, \pm 2$,  and $C_i = 0,\, \pm 1,\, \pm 2$,
%,\, \pm 3$,
being the corresponding conserved charges of the $i$th hadron: baryon number, strangeness, electric charge, and charm.

\subsection{Chemical non-equilibrium and fugacity factors}

The assumption of full chemical equilibrium can be relaxed in the heavy-ion collision applications of the HRG model.
This is usually done by introducing additional fugacity factors which regulate the absolute abundances of quarks of different flavor~\cite{Letessier:1998sz,Letessier:2005qe}, initially introduced as phenomenological parameters~\cite{Koch:1986ud,Rafelski:1991rh}.
In this case, the Boltzmann factors in Eqs.~\eqref{eq:piid}-\eqref{eq:eiid} are modified to 
\eq{
e^{\mu_i / T} \to e^{\mu_i / T} \, \gamma_q^{|q_i|} \, \gamma_S^{|S_i|} \, \gamma_C^{|C_i|} \quad \text{[or } \mu_i \to \mu_i + T \, \log(\gamma_q^{|q_i|} \, \gamma_S^{|S_i|} \, \gamma_C^{|C_i|})\text{]}, 
}
where
$|q_i|$, $|S_i|$, and $|C_i|$ correspond, respectively, to the absolute light, strange, and charm quark content of hadron $i$, 
and where $\gamma_q$, $\gamma_S$, and $\gamma_C$ are parameters which regulate deviations from chemical equilibrium in the light, strange, and charm quark sectors, respectively.
$\gamma_i = 1$ corresponds to the chemical equilibrium scenario for the corresponding quark flavor sector.

\subsection{Excluded-volume corrections}

The repulsive interactions between hadrons can be modeled by an excluded volume correction of the van der Waals type, whereby the volume available for hadrons to be created and move in is reduced by the sum of all their eigenvolumes. Such a correction was first studied in the hadronic equation of state in Refs.~\cite{Baacke:1976jv,Hagedorn:1980kb,Hagedorn:1982qh,Gorenstein:1981fa,Kapusta:1982qd}. A thermodynamically consistent procedure for a single-component gas was first formulated in Ref.~\cite{Rischke:1991ke}.

It should be noted that an excluded-volume correction is only an effective approach to treat repulsive interactions between hadrons. 
A common (and simplest) assumption is the constant eigenvolume parameter for all hadronic species~\cite{BraunMunzinger:1999qy}.
A more realistic approach is to allow the possibility for different hadrons to have different eigenvolumes~\cite{Yen:1997rv,Yen:1998pa}.
Such an approach, however, is incomplete.
For instance, it is not possible to take into account the expected differences between baryon-baryon and baryon-antibaryon interactions~\cite{Andronic:2012ut} within such a model.
Therefore, each pair of hadron species can be characterized by its own ``excluded volume'' parameter~\cite{Gorenstein:1999ce} in the most general case.

All of the above options are considered and implemented in the \thermalfist package.

\subsubsection{Diagonal EV model}

In the Diagonal EV-HRG (DEV-HRG) model~\cite{Yen:1997rv,Yen:1998pa}, each hadron is assigned an excluded-volume parameter $v_i$.
It is common to characterize the excluded-volume parameter $v_i$ with the effective hard-core radius $r_i$, by using the classical relation $v_i = (16\pi/3) \, r_i^3$.
The excluded volume correction leads to the transcendental equation for the system pressure,
\eq{\label{eq:pev}
p(T, \mu) = \sum_{i} \, p^{\rm id} (T, \mu_i^*), \qquad \mu_i^* = \mu_i - v_i \, p,
}
which is solved numerically. 
The numerical solution is obtained in \thermalfist using the Broyden's method~\cite{Broyden}.

Other thermodynamic functions are then obtained from the standard thermodynamic relations:
\eq{\label{eq:nev}
n_i(T, \mu) & \equiv \left(\frac{\partial p}{\partial \mu_i}\right)_T = \frac{n_i^{\rm id} (T, \mu_i^*)}{1 + \sum_j v_j n_j^{\rm id} (T, \mu_j^*)}, \\
\label{eq:sev}
s(T, \mu) & \equiv \left(\frac{\partial p}{\partial T}\right)_{\mu} = \frac{ \sum_i  s_i^{\rm id} (T, \mu_i^*)}{1 + \sum_j v_j n_j^{\rm id} (T, \mu_j^*)}, \\
\label{eq:eev}
\varepsilon(T, \mu) & \equiv Ts + \sum_i \mu_i \, n_i - p = \frac{\sum_i \varepsilon_i^{\rm id} (T, \mu_i^*)}{1 + \sum_j v_j n_j^{\rm id} (T, \mu_j^*)}.
}

\subsubsection{Non-diagonal EV model}
The repulsive interactions in the Non-Diagonal EV-HRG (NDEV-HRG) model~\cite{Gorenstein:1999ce,Vovchenko:2016ebv,Satarov:2016peb} are characterized by the matrix $\tilde{b}_{ij}$ of the excluded volume type parameters, which characterize the repulsive interactions for each pair of particle species.
The total pressure is partitioned into the sum of ``partial'' pressures,
\eq{
p(T,\mu) = \sum_i p_i (T,\mu),
}
which are determined by the following system of transcendental equations:
\eq{\label{eq:NDE:pi}
p_i (T,\mu) = p_i^{\rm id} (T, \mu_i^*), \qquad \mu_i^* = \mu_i - \sum_j \widetilde{b}_{ij} \, p_j, \qquad i = 1, \ldots , f.
}
The solution to Eq.~\eqref{eq:NDE:pi} at given $T$ and $\mu$ is obtained numerically in \thermalfist, using the multi-dimensional Broyden's method.

The particle number densities $n_i \equiv (\partial p / \partial \mu_i)_T$ are found as the solution to the system of linear equations
\eq{\label{eq:NDE:ni}
\sum_j [\delta_{ij} + \tilde{b}_{ji} \, n_i^{\rm id} (T, \mu_i^*)] \, n_j = n_i^{\rm id} (T, \mu_i^*), \qquad i = 1 \ldots f.
}

The entropy and energy densities are given by
\eq{\label{eq:NDE:se}
s(T, \mu) %\equiv \left(\frac{\partial p}{\partial T}\right)_{\mu} 
& = \sum_i (1 - \sum_j \tilde{b}_{ji} \, n_j) \, s^{\rm id}_i (T, \mu_i^*), \\
\varepsilon(T, \mu) %\equiv \left(\frac{\partial p}{\partial T}\right)_{\mu} 
& = \sum_i (1 - \sum_j \tilde{b}_{ji} \, n_j) \, \varepsilon^{\rm id}_i (T, \mu_i^*).
}

Parameters $\tilde{b}_{ij}$ can be chosen arbitrarily for each pair of particle species. They need not to be symmetric.
One possibility is the classical picture of a multi-component gas of hard spheres\footnote{It should be noted that the classical picture is not necessarily  valid at the nuclear scale, see~\cite{Vovchenko:2017drx}.}, where each species is characterized by the hard-core radius $r_i$ and where $\tilde{b}_{ij}$ are given by
\eq{\label{eq:NDE:clbij}
\widetilde{b}_{ij} = \frac{2\,b_{ii}\,b_{ij}}{b_{ii}+b_{jj}} \qquad \text{with} \qquad b_{ij} = \frac{2 \pi}{3} \, (r_i + r_j)^3.
}

The NDEV-HRG model reduces to the DEV-HRG model in the partial case $\widetilde{b}_{ij} \equiv v_i$.

\subsection{Quantum van der Waals model}
\label{sec:qvdw}

The presence of both, the short-range repulsive and the intermediate/long range attractive interactions between hadrons can be treated in the framework of the Quantum van der Waals (QvdW) equation~\cite{Vovchenko:2015vxa}, extended to multiple components present in a HRG~\cite{Vovchenko:2016rkn,Vovchenko:2017zpj}.
This QvdW-HRG model is defined by the following pressure function
\eq{\label{eq:vdw:p}
p(T,\mu) = \sum_i p^{\rm id}_i (T, \mu_i^*) - \sum_{i,j} \, a_{ij} \, n_i \, n_j.
}
The particle number densities, $n_i$, satisfy the system of linear equations
\eq{\label{eq:vdw:ni}
\sum_j [\delta_{ij} + \tilde{b}_{ji} \, n_i^{\rm id} (T, \mu_i^*)] \, n_j = n_i^{\rm id} (T, \mu_i^*), \qquad i = 1 \ldots f,
}
while the shifted chemical potentials, $\mu_i^*$, satisfy the system of transcendental equations
\eq{\label{eq:vdw:mu*}
\mu^*_i + \sum_j \tilde b_{ij} \, p^*_j - \sum_j (a_{ij} + a_{ji}) \, n_j = \mu_i~, \quad i = 1,\ldots,f~.
}

The pressure, $p(T,\mu)$, at a given temperature $T$ and chemical potentials $\mu$ is determined by first solving numerically the system of equations \eqref{eq:vdw:mu*} for $\mu_i^*$\footnote{Note that particle number densities $n_j$ in Eq.~\eqref{eq:vdw:mu*} are given as a function of $\{ \mu_i^* \}$ via Eq.~\eqref{eq:vdw:ni}.} and then plugging in the result into Eq.~\eqref{eq:vdw:p}.

The entropy and energy densities are given as
\eq{\label{eq:vdw:se}
s(T, \mu) 
& = \sum_i (1 - \sum_j \tilde{b}_{ji} \, n_j) \, s^{\rm id}_i (T, \mu_i^*), \\
\varepsilon(T, \mu) 
& = \sum_i (1 - \sum_j \tilde{b}_{ji} \, n_j) \, \varepsilon^{\rm id}_i (T, \mu_i^*) - \sum_{i,j} a_{ij} \,n_i \, n_j~.
}

The parameters $\tilde{b}_{ij}$ correspond to the repulsive vdW interactions, and they have the same physical meaning as in the NDEV-HRG model.
The parameters $a_{ij}$ correspond to the attractive vdW interactions\footnote{When introducing terms corresponding to the attractive vdW interactions one should take care not to double count those interactions that lead to the formation of resonances.} between hadron species $i$ and $j$, modeled in the mean-field approximation.
The QvdW-HRG model reduces to the NDEV-HRG model for the case $a_{ij} \equiv 0$.

The QvdW-HRG model permits the inclusion into the HRG model of the basic features of nuclear matter, in particular, the nuclear liquid-gas phase transition and the associated criticality. 
The critical point of nuclear matter was shown to be important for the fluctuation observables in heavy-ion collisions~\cite{Fukushima:2014lfa,Vovchenko:2016rkn,Mukherjee:2016nhb,Vovchenko:2017ayq}.

\subsection{Finite resonance widths}
\label{sec:widths}

The finite widths of the resonances may play an important role for some applications of the HRG model.
These can be taken into account in a simplified way, by adding an additional integration into Eqs.~\eqref{eq:piid}-\eqref{eq:eiid} over the resonance masses~\cite{Becattini:1995if,Wheaton:2004qb}
\eq{\label{eq:BW}
\int dk \, \to \, \int_{m_i^{\rm min}}^{m_i^{\rm max}} d m \, \rho_i(m)  \, \int dk \, ,
}
where $\rho_i(m)$ is the mass distribution for resonance $i$, taken either in the relativistic Breit-Wigner form,
\eq{
\rho_i(m) = A_i \, \frac{2 \, m \, m_i \, \Gamma_i(m)}{(m^2-m_i^2)^2 + m_i^2 \, [\Gamma_i(m)]^2}, 
}
or the nonrelativistic Breit-Wigner form,
\eq{
\rho_i(m) = A_i \, \frac{1}{(m-m_i)^2 +  [\Gamma_i(m)]^2/4}. 
}
Here $A_i$ is determined from the normalization condition $\int_{m_i^{\rm min}}^{m_i^{\rm max}} d m \, \rho_i(m) = 1$. 

There are different possibilities for the choice of $m_i^{\rm min}$ and $m_i^{\rm max}$, as well as for the mass-dependence of the width $\Gamma_i (m)$.
Two options are implemented in \thermalfist: (1) integration in the truncated interval, defined by $m_i^{\rm min} = \max(m_i - 2\Gamma_i, m_i^{\rm thr})$ and $m_i^{\rm max} = m_i + 2\Gamma_i$, with energy independent widths~\cite{Becattini:1995if,Wheaton:2004qb}, and (2) integration in the full interval, defined by $m_i^{\rm min} = m_i^{\rm thr}$ and $m_i^{\rm max} = \infty$, with energy dependent widths.
Here $m_i^{\rm thr}$ is the threshold mass.
More details about the different options and their implementation can be found in Ref.~\cite{Vovchenko:2018fmh}.

\subsection{Feeddown from resonance decays}

\begin{figure}[t]
  \centering
  \includegraphics[width=.49\textwidth]{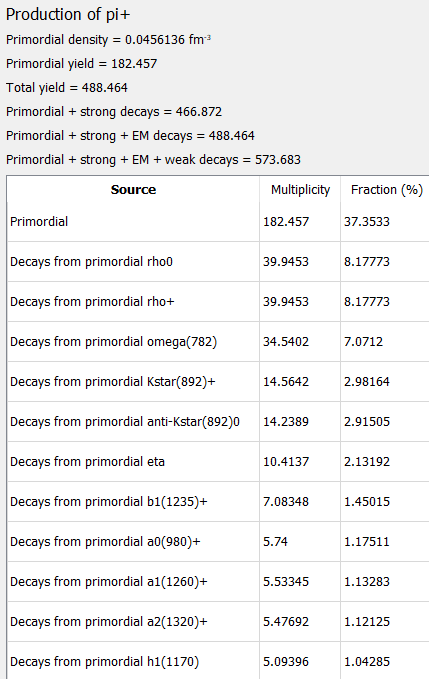}
  \includegraphics[width=.49\textwidth]{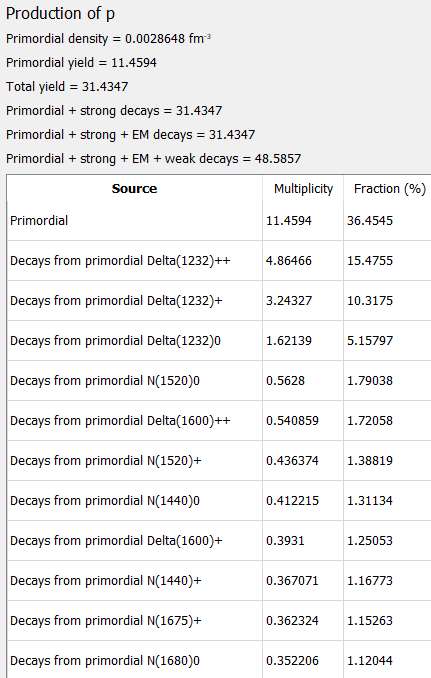}
  \caption{Screen grabs from the QtThermalFIST GUI program showing differential yield contributions to the final yields of $\pi^+$~(left panel) and protons~(right panel), evaluated in the Id-HRG model at $T = 155$~MeV, $V = 4000$~fm$^3$, and zero chemical potentials.
  }
  \label{fig:feed}
\end{figure}

Applications of the HRG model to particle production must take into account the feeding from decays of unstable particles (resonances) to the final hadron yields measured in experiments. Feeding from resonance decays gives dominant contribution in many cases, as much as 70\% of all final pions may come from resonance decays~\cite{Wheaton:2004qb}.
This is illustrated in Fig.~\ref{fig:feed}, which shows different feeddown contribution to the total yields of pions and protons, evaluated in the Id-HRG model at $T = 155$~MeV, $V = 4000$~fm$^3$, and zero chemical potentials.

The final mean multiplicity $\langle N_i\rangle$ of $i$th
particle species  is calculated in the HRG model
as a sum of the primordial mean multiplicity
$\langle N^{*}_i\rangle \equiv n_i \, V$ and resonance decay
contributions as follows
\eq{\label{eq:Ntot}
\langle N_i\rangle~ =~
\langle N^{*}_i\rangle~ +~ \sum_R \langle n_i \rangle_R \, \langle N^{*}_R\rangle~,
}
where $\langle n_i \rangle_R$ is the average number of particles
of type $i$ resulting from decay of resonance $R$. $\langle n_i \rangle_R$ includes contribution from both, the direct decays of resonance $R$ resulting in the production of hadron $i$, as well as the contributions resulting from the chain of decays via lower-mass resonances.
Note that for energy-dependent resonance widths the relation~\eqref{eq:Ntot} is modified to reflect energy dependent branching ratios, see Ref.~\cite{Vovchenko:2018fmh} for details.

The exact decay feeddown chain for each hadron yield should be matched to the relevant experimental conditions. 
By default, the hadrons which decay strongly or electromagnetically are marked unstable in \thermalfist.
It is also possible, for each hadron species, to evaluate final yields which include contributions from strong decays only, from strong and electromagnetic decays, from strong, electromagnetic and weak decays, or in accordance with manually set particle stability flags that define the feeddown.

\subsection{Correlations, fluctuations and susceptibilities}

\subsubsection{Conserved charges}

Fluctuations and correlations of conserved charges are characterized by the corresponding susceptibilities
\eq{
\label{eq:chiBSQ}
\chi_{lmnk}^{BSQC}~=~\frac{\partial^{l+m+n+k}p/T^4}{\partial(\mu_B/T)^l \,\partial(\mu_S/T)^m \,\partial(\mu_Q/T)^n\,\partial(\mu_C/T)^k}~\,.
}

These susceptibilities are useful, as they carry information about finer details of the equation of state, they
can be obtained in the lattice QCD simulations, and they can also be measured in heavy-ion collisions, although in many cases proxy observables have to be used.

In the most general case, the derivatives defining a particular $\chi_{lmnk}^{BSQC}$ can be evaluated numerically by calculating the pressure function at different values of chemical potentials.
The diagonal susceptibilities of an arbitrary conserved charge~(which can be, e.g., some linear combination of $B$, $S$, $Q$, and $C$), can be calculated within \thermalfist analytically up to the 4th order.
This analytic procedure is described in~\cite{Vovchenko:2017zpj} for the QvdW-HRG model, all other models included are obtained as partial cases of the QvdW-HRG model.

\subsubsection{Hadron yield fluctuations and probabilistic decays}
\label{sec:flucdecays}

Fluctuations and correlations of hadron numbers can also be considered. A hadron yield is not a conserved quantity, and, therefore, it cannot be calculated from first-principles using lattice QCD. On the other hand, in contrast to the conserved charges, hadron yields are quantities which are directly measured in heavy-ion collision experiments, and they are readily available for calculation in a HRG approach.

The cumulants $\chi_k^i$ of the primordial hadron number distributions, $N_i^*$, can be calculated by taking the derivatives with respect to the chemical potential $\mu_i$ of hadron specie $i$. In particular
\eq{
\chi_2^i & \equiv  \frac{\partial^2 (p/T^4)}{\partial (\mu_i/T)^2} = \frac{1}{VT^3} \, \langle (\Delta N_i^*)^2 \rangle, \\
\chi_3^i & \equiv  \frac{\partial^3 (p/T^4)}{\partial (\mu_i/T)^3} = \frac{1}{VT^3} \, \langle (\Delta N_i^*)^3 \rangle, \\
\chi_4^i & \equiv  \frac{\partial^4 (p/T^4)}{\partial (\mu_i/T)^4} = \frac{1}{VT^3} \, \langle (\Delta N_i^*)^4 \rangle_c  = \frac{1}{VT^3} \, [\langle (\Delta N_i^*)^4 \rangle - 3 \, \langle (\Delta N_i^*)^2 \rangle^2 ],
}
where $\chi_k^i$ are expressed here through the central moments of the primordial hadron number distribution.
$\chi_k^i$ are evaluated in \thermalfist in the same way as the fluctuations of conserved charges.

The two-particle correlator $\langle \Delta N_i^* \Delta N_j^* \rangle$ for the primordial hadron yields $N_i^*$ and $N_j^*$ is given by the mixed derivative
\eq{\label{eq:Nijcorr}
\frac{1}{VT^3} \, \langle \Delta N_i^* \Delta N_j^* \rangle = \frac{\partial^2 (p/T^4)}{\partial (\mu_i/T) \, \partial (\mu_j/T)}~,
}
while the higher-order correlations are obtained in a similar fashion.

The correlations and fluctuations in final hadron yields are affected by the probabilistic decays of the resonances.
The variance of the final hadron yield $N_i$ is calculated as follows~\cite{Jeon:2000wg,Begun:2006jf,Gorenstein:2007ep,Fu:2013gga}
\eq{\label{eq:dNfin}
\langle (\Delta N_i)^2 \rangle & = \langle (\Delta N_i^*)^2 \rangle + \, \sum_R \, \langle N_R^* \rangle \, \langle (\Delta n_i)^2 \rangle_R  \nonumber \\
& \quad + 2 \, \sum_R \, \langle \Delta N_i^* \Delta N_R^* \rangle \, \langle n_i\rangle_R + \sum_{R,R'} \, \langle \Delta N_R^* \, \Delta N_{R'}^* \rangle \, \langle n_i\rangle_R \, \langle n_i\rangle_{R'}.
}
Here $\langle (\Delta n_i)^2 \rangle_R$ is the variance of the number of hadrons of type $i$ which result from the chain of probabilistic decays of the resonance $R$.

The higher-order fluctuations of final hadron yields are considered in \thermalfist for the Id-HRG model only. The third and fourth cumulants in this model read~\cite{Nahrgang:2014fza}
\eq{
\langle (\Delta N_{i})^3\rangle&=\langle (\Delta N_{i}^*)^3\rangle + \sum_R \langle (\Delta N_R^*)^3\rangle \langle n_{i}\rangle_R^3
+ 3\sum_R \langle (\Delta N_R^*)^2\rangle \langle n_i\rangle_R \langle (\Delta n_{i})^2 \rangle_R \nonumber \\
 & \quad + \sum_R \langle N_R^*\rangle \langle (\Delta n_{i})^3\rangle_R\, ,\\
 \langle (\Delta N_{i})^4\rangle_c&=\langle (\Delta N_{i}^*)^4\rangle_c + \sum_R \langle (\Delta N_R^*)^4\rangle_c \langle n_{i}\rangle_R^4  + 6\sum_R \langle (\Delta N_R^*)^3\rangle \langle n_p\rangle_R^2 \langle (\Delta n_{i})^2 \rangle_R \nonumber \\
 &\quad + \sum_R \langle (\Delta N_R^*)^2\rangle \bigg[3\,\langle (\Delta n_{i})^2 \rangle_R^2 + 4\,\langle n_i\rangle_R \langle (\Delta n_{i})^3 \rangle_R \bigg] \\
 &\quad + \sum_R \langle N_R^*\rangle \langle (\Delta n_{i})^4\rangle_{R,c} \, .
}
Here $\langle (\Delta n_i)^3 \rangle_R$ and $\langle (\Delta n_{i})^4\rangle_{R,c} = \langle (\Delta n_i)^4 \rangle_R - 3 \, \langle (\Delta n_i)^2 \rangle_R^2$ are, respectively, the third and fourth cumulants of the distribution of the number of hadrons of type $i$ which result from the chain of probabilistic decays of the resonance $R$.

\subsection{Canonical ensemble}
The exact conservation of conserved charges can be enforced within the canonical thermodynamics.
The effects of exact charge conservation influence hadron yields in small systems, typically when the number of hadrons carrying a particular conserved charge is of the order of unity or smaller.
The exact charge conservation also affects strongly the fluctuations of hadron numbers, even in the thermodynamic limit.

Analytic calculations in the canonical ensemble are restricted in \thermalfist to the Id-HRG model.

\subsubsection{Full canonical ensemble}

The canonical ensemble treatment implies exact conservation of the corresponding conserved charges.
The partition function of the Id-HRG in the canonical ensemble for fixed baryon number $B$, electric charge $Q$, strangeness $S$, and charm $C$, reads~\cite{Becattini:1995if,Becattini:1997rv}
\eq{\label{eq:Z}
\mathcal{Z}(B,Q,S,C) & =
\int \limits_{-\pi}^{\pi}
  \frac{d \phi_B}{2\pi}
 \int \limits_{-\pi}^{\pi}
  \frac{d \phi_Q}{2\pi}
  \int \limits_{-\pi}^{\pi}
  \frac{d \phi_S}{2\pi}
  \int \limits_{-\pi}^{\pi}
  \frac{d \phi_C}{2\pi}~
  e^{-i \, (B \phi_B + Q \phi_Q + S \phi_S + C \phi_C)} \nonumber \\
  & \quad \times \exp\left[\sum_{j} \sum_{n = 1}^{\infty} z_{j,n} \, e^{i \, n\, (B_j \phi_B + Q_j \phi_Q + S_j \phi_S + C_j \phi_C)}\right].
}
Here
\eq{\label{eq:zjn}
z_{j,n} = (\mp 1)^{n+1} \, \gamma_q^{|q_j|} \, \gamma_S^{|S_j|} \, \gamma_C^{|C_j|} \, \frac{d_j \, V_c}{2 \pi^2} \, \frac{T \, m^2}{n^2} \, K_2\left(n \frac{m}{T} \right), 
}
with $V_c$ being the canonical correlation volume, i.e. the volume over which the exact conservation is enforced.
The mean primordial multiplicity for species $j$ is given by
\eq{
\langle N_j \rangle_{ce} = \sum_{n=1}^{\infty} \frac{Z(B-n\,B_j,Q-n\,Q_j,S-n\,S_j, C-n\,C_j)}{Z(B,Q,S,C)} \, n \, z_{j,n}.
}

Similar expressions exist for other thermodynamic functions, as well as for fluctuations and correlations of particle numbers~\cite{Begun:2006jf}.
Note that all the above relations include the effects of quantum statistics, given by the summation over $n$.
This sum is truncated at a sufficiently large $n_{\rm max}$ value in calculations.
The calculations proceed by evaluating the partition functions, given by Eq.~\eqref{eq:Z}, through a numerical integration.
For the case when both, quantum statistical effects are neglected for baryons and when there are no particles in the list with $|B_j| > 1$, the integration over $\phi_B$ is performed analytically, using the method of Ref.~\cite{Keranen:2001pr}. 
This allows to significantly speed up the calculations.

The inclusion of quantum statistics and fluctuations is a new element compared to the functionality of the presently available other open source packages.
Most packages do not include the canonical ensemble at all.
One exception is THERMUS~\cite{Wheaton:2004qb}, 
where calculations are, however, restricted to the case of Boltzmann statistics and where it is not possible to consider particles which have a baryon number $|B_j| > 1$.

It is also possible to selectively treat certain conserved charges grand-canonically, while preserving the canonical treatment for other conserved charges.
To remove the canonical treatment of a certain conserved charge, for instance the baryon charge $B$, one removes the integration over $\phi_B$ in Eq.~\eqref{eq:Z}, sets $\phi_B$ to zero in the integrand, and adds the baryon number fugacity factor into the expression~\eqref{eq:zjn} for the one-particle partition function.
The selective canonical treatment of  conserved charges can clarify the role of canonical effects for different conserved charges in a given setup.

The canonical ensemble implementation works reliably and fast when the system volume or conserved charges are sufficiently small. 
This covers the typical cases where the effects of exact charge conservation on mean hadron multiplicities are significant.

\subsubsection{Strangeness/charm canonical ensemble}

The abundances of hadrons carrying strangeness produced in intermediate-energy heavy-ion collisions 
are notably smaller than those of light flavored hadrons.
The abundances of charm hadrons are even smaller.
Therefore, in many cases it is sufficient to implement the canonical treatment of strangeness or charm only, while preserving the grand canonical treatment of baryon number and electric charge.
The canonical treatment of strangeness is achieved in the so-called strange\-ness-canonical ensemble~(SCE).
In principle, the full canonical ensemble implementation as described above provides such a functionality.
In some cases, however, the system volume can be large, and the above implementation can be inefficient.
Therefore, here we consider an alternative approach to calculate the partition functions in Eq.~\eqref{eq:Z}.
Assuming the Boltzmann statistics for strange particles, the SCE partition function is expressed as the sum over the Bessel functions~\cite{BraunMunzinger:2001as}
\eq{
\mathcal{Z}(S) = \mathcal{Z_{\rm NS}} \, \sum_{k = -\infty}^{\infty} \, \sum_{p = -\infty}^{\infty} \, a_3^p \, a_2^k \, a_1^{-2k-3p-S} \, I_k(x_2) \, I_p(x_3) \, I_{-2k-3p-S}(x_1).
}
Here $\mathcal{Z_{\rm NS}}$ is the grand-canonical partition function of the subsystem consisting of non-strange particles, and
$$
a_i = \sqrt{S_i / S_{-i}}, \qquad x_i = 2 \sqrt{S_i \, S_{-i}}, \qquad i = 1,2,3
$$
with $S_i = \sum_{j \in (S_j = i)}  \langle N_j^{\rm gce} \rangle$ being the cumulative mean multiplicity of all particle carrying strangeness $i$ calculated in the grand canonical ensemble.

The canonical treatment of charm here assumes that there are no multi-charmed particles in the particle list and that the system is net charm free, which is sufficient for most applications. The yields of charmed hadrons in the charm-canonical ensemble~(CCE) are then calculated as follows
\eq{
\langle N_i^{\rm ce} \rangle = \langle N_i^{\rm gce} \rangle \, \frac{I_1(\sum_{j \in C}  \langle N_j^{\rm gce} \rangle)}{I_0(\sum_{j \in C}  \langle N_j^{\rm gce} \rangle)}, \qquad i \in C.
}

\subsection{Thermal fits}

Perhaps the most common application of the HRG model is fitting the hadron yield data from relativistic heavy-ion collisions -- the thermal fits.
Such an approach assumes thermal and (partial) chemical equilibrium between all stable hadrons
and all resonances at the so-called ``chemical freeze-out'' stage of a heavy-ion reaction.
The HRG model fits are performed by minimizing the value
 \eq{\label{eq:xi}
 \frac{\chi^2}{N_{\rm dof}}
 ~=~\frac{1}{N_{\rm dof}}\sum_{i=1}^N\frac{\left(N_i^{\rm exp}~-~N_i^{\rm HRG}\right)^2}{\sigma_i^2}~,
}
where $N_i^{\rm exp}$ and $N_i^{\rm HRG}$ are the experimental and
calculated in the HRG hadron multiplicities, respectively; $N_{\rm
dof}$ is the number of degrees of freedom, that is the number of
the data points minus the number of fitting parameters; and
$\sigma_i^2=(\sigma_i^{syst})^2+(\sigma_i^{stat})^2$ is the sum of
the squares of the statistical and systematic experimental errors.
Note that $N_i^{\rm HRG}$ is the total hadron yield, including the resonance feeddown, calculated in accordance with Eq.~\eqref{eq:Ntot} and using the appropriate feeddown flags.
$N_i$ in Eq.~\eqref{eq:xi} can also represent a ratio of two yields.
\thermalfist employs the MINUIT2 package~\cite{James:1975dr} for the $\chi^2$ minimization procedure.

In the simplest setup, corresponding to the full chemical equilibrium in the grand canonical ensemble, there are only three fit parameters: the temperature $T$, the baryonic chemical potential $\mu_B$, and the system volume parameter $V$. The electric charge and strangeness chemical potentials $\mu_Q$ and $\mu_S$ are not fitted. Instead, at each fixed temperature $T$ and baryochemical potential $\mu_B$, the $\mu_Q$ and $\mu_S$ are determined in a unique way in order to satisfy two conservation laws given by the ``initial'' conditions: the electric-to-baryon charge ratio of $Q/B = 0.4$, and the vanishing net strangeness $S = 0$.
These two conditions are relevant if pre-freezeout radiation is neglected.
Otherwise, $\mu_S$ and/or $\mu_Q$ can also be considered as additional fit parameters.
For completeness, \thermalfist also allows to constrain the baryochemical potential $\mu_B$ to a fixed entropy-per-baryon ratio, $S/B$.

The chemical potentials disappear in the canonical ensemble formulation.
In this case, the total baryon, electric charge, strangeness, and charm numbers appear instead, these are fixed by the ``initial'' conditions.

Some modifications, such as chemical under- or over-saturation of the light, strange,  and/or charm quarks can also introduce additional parameters, $\gamma_q$, $\gamma_S$, and $\gamma_C$~(see, e.g., Refs.~\cite{Letessier:2005qe,Rafelski:2015cxa}).

\section{Monte Carlo event generator}
\label{sec:MC}

\thermalfist implements Thermal Event Generator (TEG) -- a Monte Carlo generator of hadronic microstates that correspond to a particular formulation of the HRG model in grand canonical or canonical ensemble.
The effects of radial flow are included in the framework of the blast wave model.
Monte Carlo implementation of the probabilistic decays of the primordial resonances is also provided.

The TEG is useful for heavy-ion collision applications, as well as for calculation of observables which are otherwise problematic using the analytic methods. These include e.g., effects of radial flow and momentum cuts.
Another example is the higher-order fluctuations and correlations of final hadron yields, which may include the effects of the probabilistic decays of resonances and of the residual EV/QvdW interactions between the primordial hadrons.
Monte Carlo formulation also allows to study the simultaneous effects of the EV interactions and the global charge conservation, which presently cannot be done using the analytic methods. 

The TEG assumes that hadrons stem from a thermally equilibrated source, i.e. they are emitted from a particular ``freeze-out'' hypersurface. Each point at the hypersurface is characterized by the same values of all thermal parameters of a HRG.
The generation of each event consists of three steps:
\begin{enumerate}
\item The multiplicities of all primordial hadrons are generated from the probability distribution  which corresponds to the partition function of a particular HRG model.
\item Momenta of all primordial hadrons are generated, independently for each hadron. The momentum distribution is given by the spherical or longitudinal Blast Wave model, with a kinetic freeze-out temperature parameter $T_{\rm kin}$ that can be different from the chemical freeze-out temperature $T_{\rm ch}$.
\item Probabilistic decay chain of all resonances in the system is simulated until only the stable hadrons remain (optional).
\end{enumerate}

It should be noted that the TEG is a rather simplistic event generator and has certain limitations. 
First, the present implementation is restricted to a HRG in the Boltzmann approximation, thus, the quantum statistical effects are omitted. 
Second, it is also assumed that the momenta and coordinates of the thermal hadrons at freeze-out are uncorrelated.
The validity of such approximation can be questioned if two-particle correlations at freeze-out, e.g. due to the EV interactions, are non-negligible.
Third, the modeling of decays assumes isotropic two-body and three-body decay kinematics, but provides only an approximate treatment of the isotropic many-body~(four or more) decay kinematics.
Therefore, an analysis of the observables which may be sensitive to the kinematics of many-body decays should be done with care.

\subsection{Multiplicity sampling}

The implementation of the multiplicity sampling is done according to the procedure described in Ref.~\cite{Vovchenko:2018cnf}. Here only the basic details are described.

\subsubsection{Poisson distribution}

In the simplest case one has the Id-HRG model in the grand canonical ensemble.
In this case the multiplicity distribution for each hadron species is given by the Poisson distribution, i.e. all multiplicities are described by the multi-Poisson distribution:
\eq{\label{eq:Pi}
\Pi(\{N_i\};T,V,\{\mu_Q\}) ~=~ \prod_{i=1}^f \, \frac{\langle N_i \rangle^{N_i}}{N_i!}\, e^{-\langle N_i \rangle}~,
}
where $\langle N_i \rangle \equiv \phi_i(T) \, e^{\mu_i/T} \, V$ is the mean number of hadron species $i$ in the GCE. 

In the sampling procedure, first $\langle N_i \rangle = n_i \, V$ are calculated analytically, and then $\{N_i\}$ are generated for each event independently for each hadron species $i$ from the Poisson distribution.

\subsubsection{Exact charge conservation}

In the canonical ensemble all globally conserved charges are conserved exactly.
In the Id-HRG model this condition modifies the multi-Poisson distribution as follows:
\eq{\label{eq:PiCE}
\Pi(\{N_i\};T,V,\{\mu_Q\}) ~=~ \prod_{i=1}^f \, \frac{\langle N_i \rangle^{N_i}}{N_i!}\, e^{-\langle N_i \rangle}~ \times \prod_{k=1}^q \delta(Q_k - \sum_{j=1}^f Q_k^{(j)} N_j),
}
where the index $k$ runs through all conserved charges in the system.

The implementation of exact charge conservation proceeds by combining the sampling from the multi-Poisson distribution with a rejection sampling.
To speed up the process, we use the multi-step sampling procedure described in Ref.~\cite{Becattini:2004rq}.

\subsubsection{EV/QvdW interactions}
A presence of the EV interactions leads to an appearance of a reduced volume in the partition function. In addition, the number of particles in the system is restricted from above, such that the total cumulative eigenvolume of all hadrons does not exceed the system volume.
In the QvdW-HRG model, the presence of vdW interactions leads to the following modification of the microstate probability in the GCE~(see Ref.~\cite{Vovchenko:2018cnf} for details):
\eq{\label{eq:PiEV}
\Pi_{EV}(\{N_i\};T,V,\{\mu_Q\}) & \propto \prod_{i=1}^{f} \, \frac{\left[ (V - \sum_j \tilde{b}_{ji} N_j) \,  z_i \, e^{\mu_i/T}
\right]^{N_i}}{N_i!} \, \exp\left( \sum_{j} \frac{a_{ij} N_j}{VT} N_i \right)
\nonumber \\ 
& \quad \times \Theta(\{N_i\};V), \\
\Theta(\{N_i\};V) &= \prod_{i=1}^{f} \, \theta(V - \sum_j \tilde{b}_{ji} N_j)~.
}

The Monte Carlo procedure in this case uses the importance sampling technique.
Namely, the $\{N_i\}$ multiplicities for each event are still generated from the multi-Poisson distribution, but each generated event is assigned a weight $w = \Pi_{\rm vdW} / \Pi$.
Therefore, any observable is calculated as a weighted average.
The theta function in Eq.~\eqref{eq:PiEV} is taken into account via a rejection sampling as well: all $\{N_i\}$ configurations for which $\Theta(\{N_i\};V)$ evaluates to zero are rejected.
The average multiplicities $\langle N_i \rangle$ used for the sampling from the auxiliary multi-Poisson distribution~\eqref{eq:Pi} are evaluated analytically using the GCE formulation of the QvdW-HRG model.

The multiplicity sampling for the QvdW-HRG model in the CE contains an additional rejection sampling step, as described in the previous subsection.
The DEV-HRG and NDEV-HRG models follow from the QvdW-HRG model as partial cases, as elaborated in Sec.~\ref{sec:qvdw}.

\subsection{Momentum sampling}

The thermal momenta of all the primordial hadrons are generated independently for each hadron.
A possibility of collective motion at freeze-out is included in the framework of the blast-wave model.
Two options are available: the spherically symmetric blast wave scenario~\cite{Siemens:1978pb}, appropriate for intermediate collision energies, and the cylindrically symmetric blast wave scenario~\cite{Schnedermann:1993ws}, which is more appropriate for high collision energies.

\subsubsection{Spherically symmetric blast wave model}

In the spherically symmetric blast wave model the momentum distribution is given by the Siemens-Rasmussen formula~\cite{Siemens:1978pb}:
\eq{\label{eq:SR1}
\omega_p \frac{dN}{d^3p} = \tilde{N} \, e^{-\frac{\gamma\,E}{T_{\rm kin}}} \,
\omega_p \, \left[ \left(1+\frac{T_{\rm kin}}{\gamma\,\omega_p}\right) \frac{\sinh \alpha}{\alpha} - \frac{T_{\rm kin}}{\gamma\,\omega_p} \cosh \alpha \right],
}
where $\gamma = (1-\beta^2)^{-1/2}$, $\alpha = \gamma \, \beta \, p / T_{\rm kin}$, $\beta \in [0,1)$ is the radial flow velocity parameter, and $\tilde{N}$ is the normalization constant. $T_{\rm kin}$ is the kinetic freeze-out temperature, which can be taken different from the chemical freeze-out temperature that determines the multiplicity distribution.
The momentum vector $\bvar p$ is considered in the spherical coordinate basis, i.e. $$\bvar p = (p \, \sin \theta_p \, \cos \varphi_p, p \, \sin \theta_p \, \sin \varphi_p, p \, \cos \theta_p).$$
The spherical coordinate angles $\varphi_p$ and $\theta_p$ are generated assuming the isotropic distribution, i.e. $\varphi_p$ and $\cos \theta_p$ are uniformly and independently distributed in the $[0,2\pi)$ and $[-1,1]$ intervals, respectively.
The distribution function for $p$ follows from Eq.~\eqref{eq:SR1}:
\eq{\label{eq:SR2}
\frac{dN}{dp} = 4 \pi \, p^2 \, \tilde{N} \, e^{-\frac{\gamma\,E}{T_{\rm kin}}} \, \left[ \left(1+\frac{T_{\rm kin}}{\gamma\,\omega_p}\right) \frac{\sinh \alpha}{\alpha} - \frac{T_{\rm kin}}{\gamma\,\omega_p} \cosh \alpha \right].
}

For practical purposes, it is convenient to perform the following variable change
\eq{\label{eq:pxi}
p = -p_0 \, \log \xi.
}
The new variable $\xi$ takes values in the finite interval $0 \leq \xi < 1$, which is convenient for a numerical implementation. The distribution function of $\xi$ reads
\eq{\label{eq:SR3}
\frac{dN}{d \xi} = \frac{dN}{dp} \, \left| \frac{dp}{d\xi} \right| = \frac{p_0}{\xi} \, 4 \pi \, p^2 \, \tilde{N} \, e^{-\frac{\gamma\,E}{T_{\rm kin}}} \,
 \left[ \left(1+\frac{T_{\rm kin}}{\gamma\,\omega_p}\right) \frac{\sinh \alpha}{\alpha} - \frac{T_{\rm kin}}{\gamma\,\omega_p} \cosh \alpha \right].
}

The function $d N / d \xi$ goes to zero at the edges of the interval $0 \leq \xi < 1$, and a has a maximum in-between.
Therefore, the value of $\xi$ is generated using the rejection sampling technique.
The value of the maximum of $d N / d \xi$ is determined using the ternary search.
The $p_0 = 1$~GeV value is used in the present implementation.
After the value of $\xi$ is generated, the absolute value of momentum is calculated via Eq.~\eqref{eq:pxi}.

The above procedure is repeated for all particles in the event.

\subsubsection{Cylindrically symmetric blast wave model}

The momentum distribution in the cylindrically symmetric blast wave model is given by~\cite{Schnedermann:1993ws}:
\eq{\label{eq:SSH1}
\omega_p \frac{dN}{d^3p} & = \tilde{N} \, m_T \, \int_{-\eta_{\rm max}}^{\eta_{\rm max}} 
d \eta \cosh (y - \eta) \, \int_0^1 \tilde{r} \, d \tilde{r}  \nonumber \\
& \quad \times \exp\left[ - \frac{m_T \, \cosh \rho \, \cosh (y - \eta)}{T_{\rm kin}} \right] 
\, I_0 \left(\frac{p_T \sinh \rho}{T_{\rm kin}}\right),
}
where $m_T = \sqrt{p_T^2 + m^2}$ is the transverse mass, $y = \displaystyle \frac{1}{2} \log \frac{\omega_p - p_z}{\omega_p + p_z}$ is the longitudinal rapidity, $\rho = \tanh^{-1} \beta_r$, and $\beta_r = \beta_s \tilde{r}^n$ is the transverse flow velocity profile, $\eta_{\rm max}$ is the longitudinal rapidity cutoff. 
$\beta_s$ is the transverse flow velocity at the surface.
The mean transverse flow velocity is $\langle \beta_T \rangle = \frac{2}{2+n} \, \beta_s$.
The particle three-momentum is parameterized as $\bvar p = (p_T \, \cos \varphi_p, p_T \sin \varphi_p, m_T \sinh y)$.

Due to the azimuthal symmetry, the azimuthal angle $\varphi_p$ is distributed uniformly in the interval $[0,2\pi)$, and it is independent of $p_T$ and $y$ . 

The distribution function for the transverse momentum $p_T$ reduces to~\cite{Schnedermann:1993ws}:
\eq{\label{eq:SSH2}
\frac{dN}{d p_T} \propto p_T \, m_T \, \int_0^1 \tilde{r} \, d \tilde{r} \,  
I_0 \left(\frac{p_T \sinh \rho}{T_{\rm kin}}\right) \, K_1 \left(\frac{m_T \cosh \rho}{T_{\rm kin}}\right).
}
The $p_T$ value is generated 
using the change of variable $p_T = -p_0 \, \log \xi$, and the rejection sampling technique for the generation of $\xi$, using the same method as described in the previous subsection for the Siemens-Rasmussen formula.

The rapidity distribution at a fixed value of $p_T$ follows from Eq.~\eqref{eq:SSH1}:
\eq{\label{eq:SSH3}
\frac{dN}{dy} (y~|~p_T) & \propto \int_{-\eta_{\rm max}}^{\eta_{\rm max}} 
d \eta \cosh (y - \eta) \, \int_0^1 \tilde{r} \, d \tilde{r}  \nonumber \\
& \quad \times \exp\left[ - \frac{m_T \, \cosh \rho \, \cosh (y - \eta)}{T_{\rm kin}} \right] 
\, I_0 \left(\frac{p_T \sinh \rho}{T_{\rm kin}}\right).
}
The value of $y$ is generated from Eq.~\eqref{eq:SSH3} using the rejection sampling technique.

\section{\thermalfist structure and implementation}
\label{sec:structure}

\thermalfist is implemented as a C++ library of classes and functions.
The applications of the package proceed by linking the \texttt{ThermalFIST} library within a C++ program (macro), which performs the needed task.

The library documentation is available online~\cite{FIST-doc}.
Here only the basic structure is described.

\subsection{Base classes}

\subsubsection{\texttt{IdealGasFunctions} module}

Calculation of thermodynamic properties of an ideal gas in the GCE forms the basis of all HRG calculations in \thermalfist.
These are given as functions of the temperature $T$, the chemical potential $\mu$, the particle mass $m$, the internal degeneracy factor $d$, and the statistics $\eta$. Here $\eta = +1$ corresponds to the Fermi-Dirac statistics, $\eta = -1$ corresponds to the Bose-Einstein statistics, and $\eta = 0$ corresponds to the Maxwell-Boltzmann statistics.

These functions are implemented in the \texttt{IdealGasFunctions} module, and include pressure, particle number density, energy density, entropy density, scalar density, and the leading four particle number susceptibilities $\chi_k$.

The calculations within the Maxwell-Boltzmann statistics~($\eta = 0$) use known analytical expressions involving the modified Bessel functions. For example, the pressure reads
\eq{\label{eq:pMB}
p^{\rm id}_{MB} (T,\mu) = \frac{d \, m^2\, T^2}{2\pi^2} \, K_2(m/T).
}
Similar expressions exist for all other ideal gas functions.

There are two options for the numerical calculations of the ideal quantum gas functions~($\eta = \pm 1$): (i) using the cluster expansion technique, (ii) using the numerical integration with Gauss-Laguerre quadratures (the default method).

\paragraph{Cluster expansion}

The first method uses the cluster expansion of an ideal quantum gas. The calculations proceed by calculating the truncated cluster expansion, e.g., the pressure is calculated as
\eq{
p^{\rm id}_{q} (T,\mu) \simeq \frac{d \, m^2\, T^2}{2\pi^2} \sum_{k=1}^{k_{\rm max}} \frac{(-\eta)^{k+1}}{k^2} \, e^{k\, \mu / T} \, K_2(k \, m/T),
}
similar expressions are used for other thermodynamic functions.
The value of $k_{\rm max}$ can be regulated, it should be sufficiently large for accurate calculations. Setting $k_{\rm max} = 1$ one recovers the Maxwell-Boltzmann statistics. Note that cluster expansion formally diverges if $\mu > m$. Therefore, this method should not be used for calculating the Fermi-Dirac functions for $\mu > m$, and extra care taken for the case $\mu \lesssim m$.

\paragraph{Numerical integration}

The numerical integration method proceeds by evaluating the integrals numerically, with the help of quadratures.
First, the integrals are written in the dimensionless form, 
\eq{\label{eq:pdimless}
p^{\rm id}_{q} (T,\mu) & = \frac{d}{6\pi^2} \, \int_{0}^{\infty} \, dk \, \frac{k^4}{\sqrt{m^2 + k^2}} \, \left\{ \exp\left( \frac{\sqrt{m^2+k^2}-\mu}{T}\right) + \eta \right\}^{-1} \nonumber \\
& = \frac{d \, T^4}{6 \pi^2} \, \int_{0}^{\infty} \, d \tilde{k} \, \frac{\tilde{k}^4}{\sqrt{\tilde{m}^2 + \tilde{k}^2}} \, \left\{ \exp\left( \sqrt{\tilde{m}^2 + \tilde{k}^2} - \tilde{\mu} \right) + \eta \right\}^{-1},
}
where $\tilde{k} = k / T$, $\tilde{m} = m / T$, and $\tilde{\mu} = \mu / T$.

For $\mu < m$, the calculations proceed by applying the 32-point Gauss-Laguerre quadrature to the dimensionless integral in Eq.~\eqref{eq:pdimless}.

For $\mu > m$, the Bose-Einstein integrals are divergent.
The Fermi-Dirac integrals, however, are convergent and can be computed. The numerical accuracy of the above-described method, however, is usually unsatisfactory in such a case.
Therefore, for calculating the Fermi-Dirac integrals at $\mu > m$, Eq.~\eqref{eq:pdimless} is rewritten as follows~\cite{LL,Satarov:2009zx}:
\eq{\label{eq:FermiLargeMu}
p^{\rm id}_{\rm FD} (T,\mu)
& \stackrel{\mu>m}{=} -\frac{d \, T^4}{6 \pi^2} \, \int_{0}^{\tilde{p}_F} \, d \tilde{k} \, \frac{\tilde{k}^4}{\sqrt{\tilde{m}^2 + \tilde{k}^2}} \, \left\{ \exp\left( \tilde{\mu} - \sqrt{\tilde{m}^2 + \tilde{k}^2} \right) + 1 \right\}^{-1}
\nonumber \\
& \quad + 
\frac{d \, T^4}{6 \pi^2} \, \int_{\tilde{p}_F}^{\infty} \, d \tilde{k} \, \frac{\tilde{k}^4}{\sqrt{\tilde{m}^2 + \tilde{k}^2}} \, \left\{ \exp\left( \sqrt{\tilde{m}^2 + \tilde{k}^2} - \tilde{\mu} \right) + 1 \right\}^{-1}
\nonumber \\
& \quad + \frac{d \, T^4}{6 \pi^2} \left[ \tilde{\mu} \, \tilde{p}_F^3 - \frac{3}{4} \, \tilde{p}_F^4 \, \psi(\tilde{m} / \tilde{p}_F) \right].
}
Here $\tilde{p}_F = \sqrt{\tilde{\mu}^2 - \tilde{m}^2}$ and 
\eq{
\psi(x) = \left(1 + \frac{x^2}{2}\right) \sqrt{1+x^2} - \frac{x^4}{2} \ln\left(\frac{1+\sqrt{1+x^2}}{x}\right).
}
The first integral in Eq.~\eqref{eq:FermiLargeMu} is evaluated using the 32-point Gauss-Legendre quadrature, while the second integral in Eq.~\eqref{eq:FermiLargeMu} is evaluated using the 32-point Gauss-Laguerre quadrature.

Expressions similar to Eq.~\eqref{eq:FermiLargeMu} are obtained for other quantities using the standard thermodynamic relations.

\subsubsection{\texttt{ThermalModelParameters}}

The HRG model parameters are kept in the \texttt{ThermalModelParameters} structure.
Each implementation of the HRG model has this structure, which contains the values of thermal parameters used in calculations.
List of all parameters is presented in Table~\ref{tab:params}.

\begin{table}
 \caption{List of HRG model parameters contained in the \texttt{ThermalModelParameters} structure. } %title
 \centering                                                 %centering table
 \begin{tabular}{ccc}                                   %columns {format}
 \hline
 \hline
 Parameter & Unit & Notes\\
 \hline
 Temperature ($T$) & GeV & -- \\
 Baryon chemical potential ($\mu_B$) & GeV & Not in CE\\
 Electric chemical potential ($\mu_Q$) & GeV & Not in CE \\
 Strangeness chemical potential ($\mu_S$) & GeV & Not in CE/SCE \\
 Charm chemical potential ($\mu_C$) & GeV & Not in CE/SCE/CCE \\
 $\gamma_q$ & -- & -- \\
 $\gamma_S$ & -- & -- \\
 $\gamma_C$ & -- & -- \\
 Volume ($V$) & fm$^3$ & -- \\
 Canonical  volume ($V_c$) & fm$^3$ & In CE, SCE, CCE \\
 Baryon charge ($B$) & -- & CE only \\
 Electric charge ($Q$) & -- & CE only \\
 Strangeness ($S$) & -- & CE, SCE \\
 Charm ($C$) & -- & CE, CCE \\
\hline
\hline
 \end{tabular}
\label{tab:params}
\end{table}

\subsubsection{\texttt{ThermalParticle} class}

The \texttt{ThermalParticle} class contains all information about the properties of a particular hadron species in the hadron list.
This includes the Particle Data Group (PDG) code, the mass, the degeneracy, the type of quantum statistics, the quantum numbers, absolute quark contents, and, if applicable, the decay channels and the width.

The class also provides various methods. 
The calculation of the ideal gas functions is performed through the \texttt{Density} method.
These calculations take into account a possible additional integration due to a finite width. 
Other methods provide setting of various options,
including whether the integrals are evaluated using the cluster expansion or the numerical integration,
the number of terms in the cluster expansion~(if this method is used), the shape and the prescription used to treat the finite resonance widths.
More details can be found in the online documentation~\cite{FIST-doc}, in particular in the annotated \texttt{ThermalParticle.h} source file.

\subsubsection{\texttt{ThermalParticleSystem} class}

The \texttt{ThermalParticleSystem} class contains the list of particles to be used in calculations, represented by a vector of \texttt{ThermalParticle} objects, one per each particle species.

The particle list is usually provided from an external file.  The \texttt{LoadList} method of the class loads the particle list from an external file.
The external file which contains the particle list has a format of a table, where each row correspond to a single particle specie.
Each column in the table corresponds to a particular hadron property. 
List of all columns is provided in Table~\ref{tab:input}.

\begin{table}
 \caption{List of hadron properties contained in an input file. } %title
 \centering                                                 %centering table
 \begin{tabular}{ccc}                                   %columns {format}
 \hline
 \hline
 %Parameters & Values \\
 %\hline
 Column & Value type & Description\\
 \hline
 PDG ID & \texttt{long long} & PDG code \\
 Name & \texttt{string} & Particle name\\
 Stability flag & \texttt{bool} & Stability w.r.t decays \\
 Mass [GeV] & \texttt{double} & Particle mass \\
 Degeneracy &  \texttt{double} & Internal degeneracy factor \\
 Statistics &  \texttt{int} & Fermi~(+1), Bose~(-1), Boltzmann~(0) \\
 $B$ & \texttt{int} & Baryon number \\
 $Q$ & \texttt{int} & Electric charge \\
 $S$ & \texttt{int} & Strangeness \\
 $C$ & \texttt{int} & Charm \\
 $n_{|S|}$ & \texttt{double} & Absolute strange quark content \\
 $n_{|C|}$ & \texttt{double} & Absolute charm quark content \\
 Width~[GeV] & \texttt{double} & Resonance width \\
 Threshold~[GeV] & \texttt{double} &  Threshold mass \\
\hline
\hline
 \end{tabular}
\label{tab:input}
\end{table}

\subsubsection{\texttt{ThermalModel} classes}

Implementation of the different HRG models is contained in classes starting with the prefix \texttt{ThermalModel}.

\paragraph{\texttt{ThermalModelBase}}

\texttt{ThermalModelBase} is the base class for all \texttt{ThermalModel} classes.
This is an abstract class containing fields and methods common for an arbitrary variant of HRG model. The exact implementation of some of these methods can be different in different versions of HRG, and is given in the corresponding derived classes.
The \texttt{ThermalModelBase} class instance contains a pointer to a \texttt{ThermalParticleSystem} object containing the particle list, and a \texttt{ThermalModelParameters} structure with all the thermal model parameters. 
Both are provided during the creation of the \texttt{ThermalModelBase} instance.
Some of the important methods of the %\texttt{ThermalModelBase} 
class include:
\begin{itemize}
\item Methods to set the various thermal model parameters. The list is given in Table~\ref{tab:setparams}.
\item \texttt{SetUseWidth} method to specify the treatment of finite resonance widths.
\item \texttt{ConstrainMuB(bool)}, \texttt{ConstrainMuQ(bool)}, \texttt{ConstrainMuS(bool)},\\ and \texttt{ConstrainMuC(bool)} methods to define whether the values of the corresponding chemical potentials should be constrained by, respectively, a fixed entropy per baryon ratio $S/B$, electric-to-baryon charge ratio $Q/B$, zero strangeness and zero charm. 
\item %\texttt{FixParameters()} 
\texttt{ConstrainChemicalPotentials()}
procedure constrains, if required, the chemical potentials $\mu_B$, $\mu_Q$, $\mu_S$, and/or $\mu_C$.
\item \texttt{CalculatePrimordialDensities()} procedure calculates the primordial hadronic densities for current values of the thermal parameters. This is the main method were most calculations take place.
\item \texttt{CalculateFeeddown()} procedure calculates various feeddown contributions to final yields (densities) of hadrons after the primordial densities were computed.
\item \texttt{double GetDensity(int PDGID, Feeddown::Type feeddown)} returns the calculated density for a particle with the given PDG code and feeddown. The \texttt{feeddown} parameter can be 0 (primordial density), 1~(with feeddown according to the stability flags),
2 (with strong/electromagnetic/weak decay feeddown),
3 (with strong/electromagnetic decay feeddown), and 4 (with strong decay feeddown).
This method should only be called after the primordial and final densities were calculated.
\end{itemize}

\begin{table}
 \caption{List of the \texttt{ThermalModelBase} methods to set HRG parameters. } %title
 \centering                                                 %centering table
 \begin{tabular}{ccc}                                   %columns {format}
 \hline
 \hline
 Method & Parameter & Notes\\
 \hline
 \texttt{SetTemperature} & $T$ & -- \\
 \texttt{SetBaryonChemicalPotential} & $\mu_B$ & -- \\
 \texttt{SetElectricChemicalPotential} & $\mu_Q$ & -- \\
 \texttt{SetStrangenessChemicalPotential} & $\mu_S$ & -- \\
 \texttt{SetCharmChemicalPotential} & $\mu_Q$ & -- \\
 \texttt{SetGammaq} & $\gamma_q$ & -- \\
 \texttt{SetGammaS} & $\gamma_S$ & -- \\
 \texttt{SetGammaC} & $\gamma_C$ & -- \\
 \texttt{SetVolume} & $V$        & -- \\
 \texttt{SetVolumeRadius} & $R$  & Sets $V = (4/3) \pi R^3$ \\
 \texttt{SetCanonicalVolume} & $V_c$ & CE, SCE, CCE \\
 \texttt{SetCanonicalVolumeRadius} & $R_c$  & Sets $V_c = (4/3) \pi R_c^3$ \\
 \texttt{SetBaryonCharge}   & $B$ & For CE only \\
 \texttt{SetElectricCharge} & $Q$ & For CE only \\
 \texttt{SetStrangeness}    & $S$ & CE, SCE \\
 \texttt{SetCharm}          & $C$ & CE, CCE \\
\hline
\hline
 \end{tabular}
\label{tab:setparams}
\end{table}

Classes which implement the different variants of the HRG model inherit from the \texttt{ThermalModelBase} class and are listed in Table~\ref{tab:classes}.
More details about the usage of the \texttt{ThermalModel} classes can be found in the corresponding source files and in the annotated sample routines in the \texttt{src/examples} folder.

\begin{table}
 \caption{List of classes inheriting from the \texttt{ThermalModelBase} class and the corresponding HRG model. } %title
 \centering                                                 %centering table
 \begin{tabular}{ccc}                                   %columns {format}
 \hline
 \hline
 Class & HRG model variant & Ensemble\\
 \hline
 \texttt{ThermalModelIdeal} & Id-HRG & GCE \\
 \texttt{ThermalModelDiagonalEV} & Diagonal EV-HRG & GCE \\
 \texttt{ThermalModelCrosstermsEV} & Non-diagonal EV-HRG & GCE \\
 \texttt{ThermalModelVDW} & QvdW-HRG & GCE \\
 \texttt{ThermalModelCanonical} & Id-HRG & CE/GCE \\
 \texttt{ThermalModelCanonicalStrangeness} & Id-HRG & SCE \\
 \texttt{ThermalModelCanonicalCharm} & Id-HRG & CCE \\
\hline
\hline
 \end{tabular}
\label{tab:classes}
\end{table}

\subsection{\texttt{ThermalModelFit} class}

The \texttt{ThermalModelFit} class implements routines related to the thermal fitting of hadron yields and/or yield ratios.
The class instance contains a pointer to the \texttt{ThermalModelBase} class instance, provided on object initialization, as well as the set of parameters for fitting, and the data to fit. 

The list of all possible fit parameters is shown in Table~\ref{tab:fitparams}.
If the parameters $\mu_Q$, $\mu_S$, or $\mu_C$ are not fitted, then they are constrained by the corresponding $Q/B$ ratio and strangeness/charm neutrality conditions.
Obviously, (some of)~these parameters are not used in the fit also if CE/SCE/CCE ensemble is employed where these parameters do not appear at all.

\begin{table}
 \caption{List of fit parameters in a thermal model fit, their default values and fit flags} %title
 \centering                                                 %centering table
 \begin{tabular}{ccc}                                   %columns {format}
 \hline
 \hline
 Parameter & \multicolumn{2}{c}{Default} \\
 \hline
 & value & fit flag \\
 \hline
 $T$ & 0.150~GeV & true \\
 $\mu_B$ & 0 & true \\
 $\mu_Q$ & 0 & false \\
 $\mu_S$ & 0 & false \\
 $\mu_C$ & 0 & false \\
 $\gamma_q$ & 1 & false \\
 $\gamma_S$ & 1 & false \\
 $\gamma_C$ & 1 & false \\
 $R$ & 8~fm & true \\
 $R_c$  & 8~fm & true \\
\hline
\hline
 \end{tabular}
\label{tab:fitparams}
\end{table}

The data points to fit can be either read from an external file with the \texttt{loadExpDataFromFile(string filename)} method, or added manually one-by-one through the \texttt{AddQuantity} method.
See the annotated\\ 
\href{https://github.com/vlvovch/Thermal-FIST/blob/master/src/examples/BagModelFit/BagModelFit.cpp}{\texttt{src/examples/BagModelFit/BagModelFit.cpp}} macro for an example of using the \texttt{ThermalModelFit} class.

\subsection{Event generator}

The thermal event generators with momentum distribution given by the spherical or longitudinally symmetric Blast-Wave model are implemented in  \texttt{SphericalBlastWaveEventGenerator} and \texttt{LongitudinalBlastWaveEventGenerator} classes, respectively.
The underlying thermal model configuration, as well as the momentum spectrum parameters described in Sec.~\ref{sec:MC}, are provided through the constructor. 
A new event can be generated with the \texttt{GetEvent()} method and
written into file with the \texttt{GetEvent().writeToFile()} method.

\subsection{Graphical user interface (GUI)}

The package contains \texttt{QtThermalFIST} module -- a GUI frontend for the \thermalfist library, where some general purpose HRG model tasks can be performed in a convenient way.
The GUI is written using the cross-platform framework Qt5.
\texttt{QtThermalFIST} consists of five tabs.

In the \texttt{Thermal model} tab a comprehensive HRG model calculation can be performed at given values of thermal parameters.
It is possible to analyze the primordial and total yields of all hadron species, particle number fluctuations, and the equation of state properties.
Calculations can be performed in the grand canonical or canonical ensemble, within the Id-HRG, DEV-HRG, NDEV-HRG, or QvdW-HRG model, using a number of common eigenvolume parametrizations.
Inclusion of quantum statistics is optional. 
There is also a possibility to constrain electric, strangeness, and/or charm chemical potentials from conservation laws, and to specify different prescriptions for the treatment of finite resonance widths.

Thermal fitting of the hadron production data is performed in the \texttt{Thermal fits} tab. It is possible to specify which parameters are fitted~(and in which range) and which are fixed. The experimental data can be loaded from an external file. Some samples are provided in the \href{https://github.com/vlvovch/Thermal-FIST/blob/master/input/data}{\texttt{input/data}} folder. Possibility to input the data directly inside the GUI is provided as well. It is also possible to view the fit results in a form of thermal fit plots. Another useful feature is calculation of the $\chi^2$ profiles.

The \texttt{Equation of state} tab offers a possibility to study the temperature dependence at a fixed $\mu_B$ of some common equation of state observables, conserved charges susceptibilities, and particle number densties (primordial or with feeddown). It is also possible to consider ratios of any pair of these observables. At finite $\mu_B$ the chemical potentials $\mu_Q$ and $\mu_S$ can be fixed from a fixed $Q/B$ ratio and zero net strangeness, or set to zero otherwise. 
For a number of observables at $\mu_B = 0$ the published lattice QCD data of the Wuppertal-Budapest and/or HotQCD collaborations is plotted along with the calculation results for convenience.

The \texttt{Event generator} tab can be used for generating events with the TEG.
The GUI program shows histograms with momentum distributions of the different generated hadrons. It is also possible here to write the generated events to a specified file.

The \texttt{Particle list editor} tab provides a user-friendly editor of the particle list. The edited list can be saved to a file and/or used on the fly in HRG model calculations in the other tabs.

In general, most of the options in \texttt{QtThermalFIST} are self-explanatory, a more detailed description of the usage can be found in the \href{https://github.com/vlvovch/Thermal-FIST/blob/master/docs/quickstart.md}{Quick Start Guide} which comes with the package.

\section{Installation}
\label{sec:install}

The current version of the source code of \thermalfist can be obtained from \href{https://github.com/vlvovch/Thermal-FIST}{https://github.com/vlvovch/Thermal-FIST}

The package is platform-independent. 
The preferred method to configure the package is to use \texttt{{\bf cmake}}.
An example for a Linux system:
\begin{lstlisting}[language=bash]
# Clone the repository from GitHub
git clone https://github.com/vlvovch/Thermal-FIST.git
cd Thermal-FIST

# Create a build directory, configure the project with cmake 
# and build with make 
mkdir build
cd build
cmake ../
make

# Run the GUI frontend
./bin/QtThermalFIST

# Run the test calculations from the paper
./bin/examples/cpc1HRGTDep
./bin/examples/cpc2chi2
./bin/examples/cpc3chi2NEQ
./bin/examples/cpc4mcHRG
\end{lstlisting}
The above commands will build the \thermalfist package in the \texttt{build} directory within the root folder of the package, run the GUI frontend program, QtThermalFIST, if it was built, and various test calculations which are presented below. 
More specifically, the library will be located in \texttt{build/lib} directory, the QtThermalFIST program in \texttt{build/bin}, and the sample macros for using the library in \texttt{build/bin/examples}.
Note that in order to build the QtThermalFIST GUI one may need to install first the freely available Qt5 framework~\cite{Qt5}.

Automated tasks can be performed by writing C++ macros which link to the \thermalfist library and perform the necessary calculations. 
Sample macros in \href{https://github.com/vlvovch/Thermal-FIST/tree/master/src/examples}{\texttt{src/examples}} can be used as templates.

\section{Test results}
\label{sec:results}

\subsection{Thermodynamic properties at zero chemical potential}

Here we consider the temperature dependence of thermodynamic functions, namely the pressure and the energy density, calculated at $\mu = 0$ within different variants of the HRG model.
The \thermalfist results are compared to the calculations performed using the \textsc{THERMUS-2.3} package.
In order to ensure a consistent comparison, the \thermalfist calculations here employ the hadron list from \textsc{THERMUS-2.3}.
Quantum statistics and the energy independent Breit-Wigner scheme for resonance widths are used in both codes.

Three different variants of the HRG model are considered.
The first variant is the standard Id-HRG model.
The second variant is the EV-HRG model with a constant radius parameter $r = 0.3$~fm assigned to all hadron species. Equation of state within this particular EV-HRG model was earlier considered in Refs.~\cite{Andronic:2012ut,Vovchenko:2014pka}.
The third variant is the QvdW-HRG model, which includes vdW interactions between baryon-baryon and antibaryon-antibaryon pairs, with vdW parameters $a = 329$~MeV\,fm$^3$ and $b = 3.42$~fm$^3$, common for all pairs. 
This model was formulated in Ref.~\cite{Vovchenko:2016rkn}.

\begin{figure}[t]
  \centering
  \includegraphics[width=.49\textwidth]{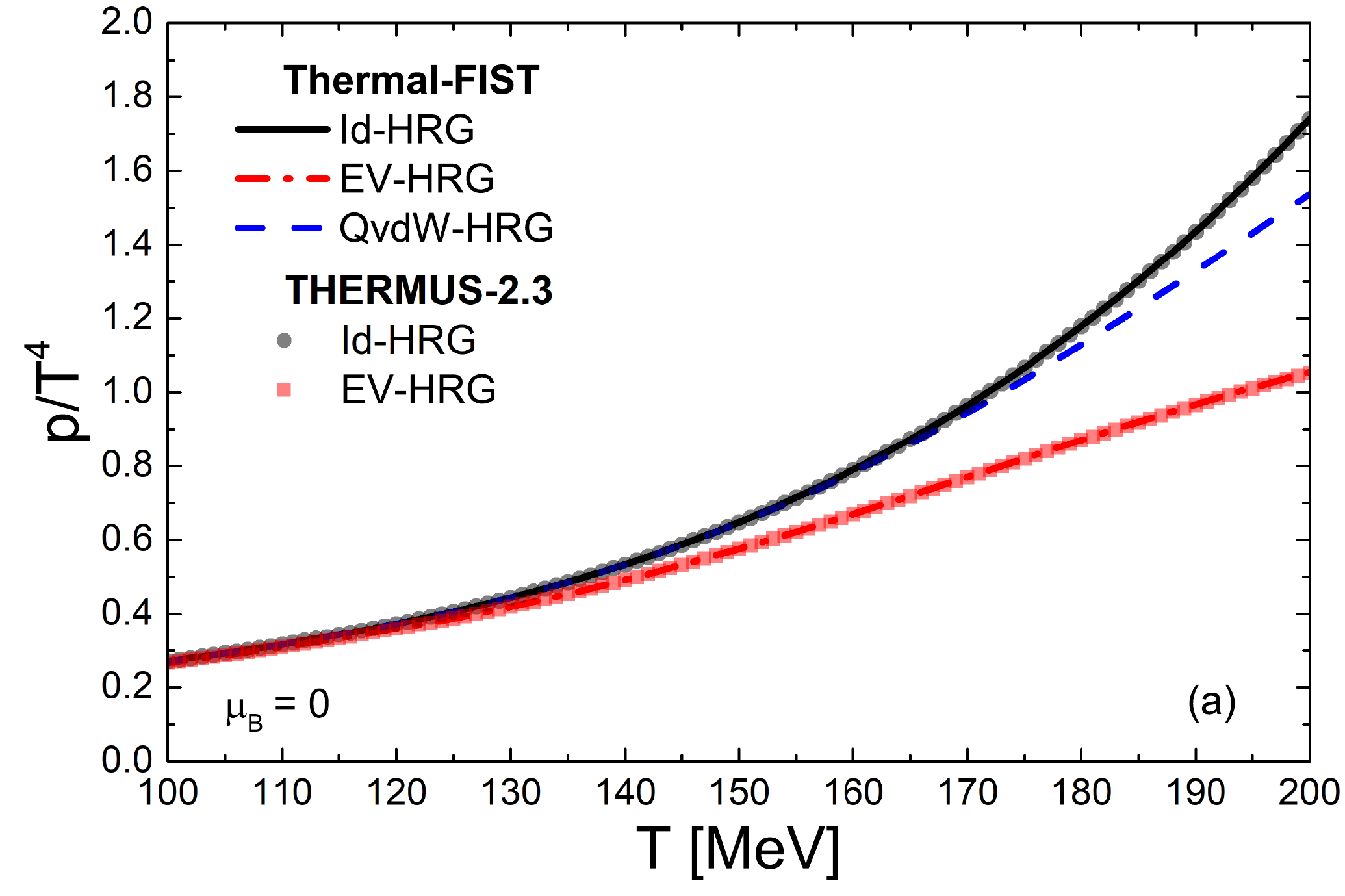}
  \includegraphics[width=.49\textwidth]{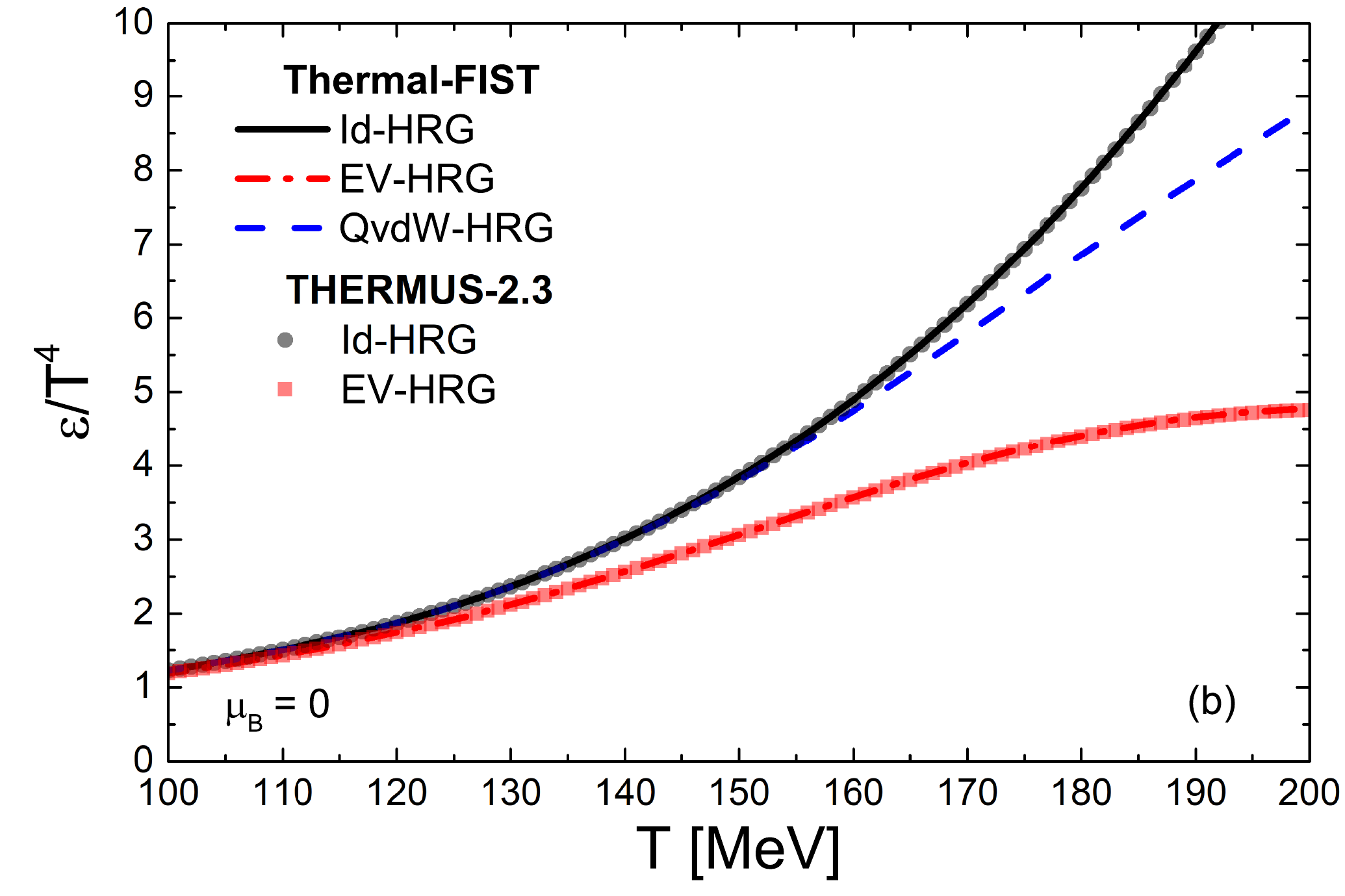}
  \caption{Temperature dependence of (a) scaled pressure $p/T^4$ and (b) scaled energy density $\varepsilon/T^4$, calculated within the Id-HRG model (solid black lines and circles), the EV-HRG model with constant radius parameter $r=0.3$~fm for all hadron species~(dash-dotted red lines and squares), and the QvdW-HRG model with a vdW interaction between baryons~(dashed blue lines). 
  The lines correspond to calculations within \thermalfist, while the symbols depict \textsc{THERMUS-2.3} calculations.
  Hadron list from \textsc{THERMUS-2.3} is used in \thermalfist calculations.
  }
  \label{fig:thermodynamics}
\end{figure}

The temperature dependence of the scaled pressure $p/T^4$ and the scaled energy density $\varepsilon / T^4$ calculated for the three different HRG models described above are depicted in Fig.~\ref{fig:thermodynamics} by lines~(\textsc{FIST}) and symbols~(\textsc{THERMUS-2.3}).
One can see that the EV/vdW effects kick in at higher temperatures, where hadronic densities become large.
The \thermalfist and \textsc{THERMUS-2.3} results for Id-HRG and EV-HRG models are quantitatively consistent with each other.
It is unfortunately not possible to present such a comparison for the QvdW-HRG model since \textsc{THERMUS-2.3} does not contain an implementation of this model.

The annotated macro, which performs the above calculations, can be found in the following location: \href{https://github.com/vlvovch/Thermal-FIST/blob/master/src/examples/cpc/cpc1-HRG-TDep.cpp}{\texttt{src/examples/cpc/cpc1-HRG-TDep.cpp}}

\subsection{Thermal fits to heavy-ion hadron yield data}

\subsubsection{Excluded volume/van der Waals effects}

In this subsection we consider the thermal fits to hadron yield data.
It is illustrated how different EV/vdW parametrizations influence the fits.
For testing purposes, we consider here those EV/vdW configurations which have a strong influence on thermal fits.
In particular, we redo here some of the calculations which were presented in Ref.~\cite{Vovchenko:2015cbk}.
We take the hadron yield data measured by the ALICE collaboration in 0-5\% most central Pb-Pb collisions at $\sqrt{s_{NN}} = 2.76$~TeV.
The actual data used in the fits can be found in Table 1 of Ref.~\cite{Vovchenko:2015cbk}.
We assume $\mu = 0$, and consider the temperature dependence of $\chi^2$ of the fit.
To calculate this temperature dependence, we minimize $\chi^2$ at each temperature by fitting the system volume parameter.

Four different variants of the HRG model are considered here.
\begin{enumerate}
\item The ideal HRG model.
\item The DEV-HRG model with mass-proportional eigenvolumes, $v_i = v_p \, (m_i / m_p)$, where $v_p = (16/3) \pi r_p^3$ with $r_p = 0.5$~fm is the assumed proton eigenvolume, and $m_p = 0.938$~GeV/$c^2$ is the proton mass. This is the bag model HRG considered in Ref.~\cite{Vovchenko:2015cbk}.
\item The DEV-HRG model with point-like mesons, and finite-sized (anti)baryons with $v_B = (16/3) \pi r_B^3$, $r_B = 0.3$~fm. 
This HRG model was considered in Refs.~\cite{Vovchenko:2015cbk,Andronic:2012ut}.
\item The QvdW-HRG model which includes the QvdW interactions for baryon-baryon and antibaryon-antibaryon pairs, with QvdW parameters $a = 329$~MeV\,fm$^3$ and $b = 3.42$~fm$^3$, common for all pairs. 
This model was formulated in Ref.~\cite{Vovchenko:2016rkn}.
\end{enumerate}

As in the previous example, the calculations are performed using both \thermalfist and \textsc{THERMUS-2.3}.
Therefore, \thermalfist calculations use the hadron list from \textsc{THERMUS-2.3}.
Quantum statistics and the energy independent Breit-Wigner modeling of finite resonance widths are used in both codes.
In order to correctly calculate the feeddown contribution to the yield of the unstable $\phi$ meson, we apply the fix~\cite{thermus-decays-fix} to THERMUS-2.3.

\begin{figure}[t]
  \centering
  \includegraphics[width=.70\textwidth]{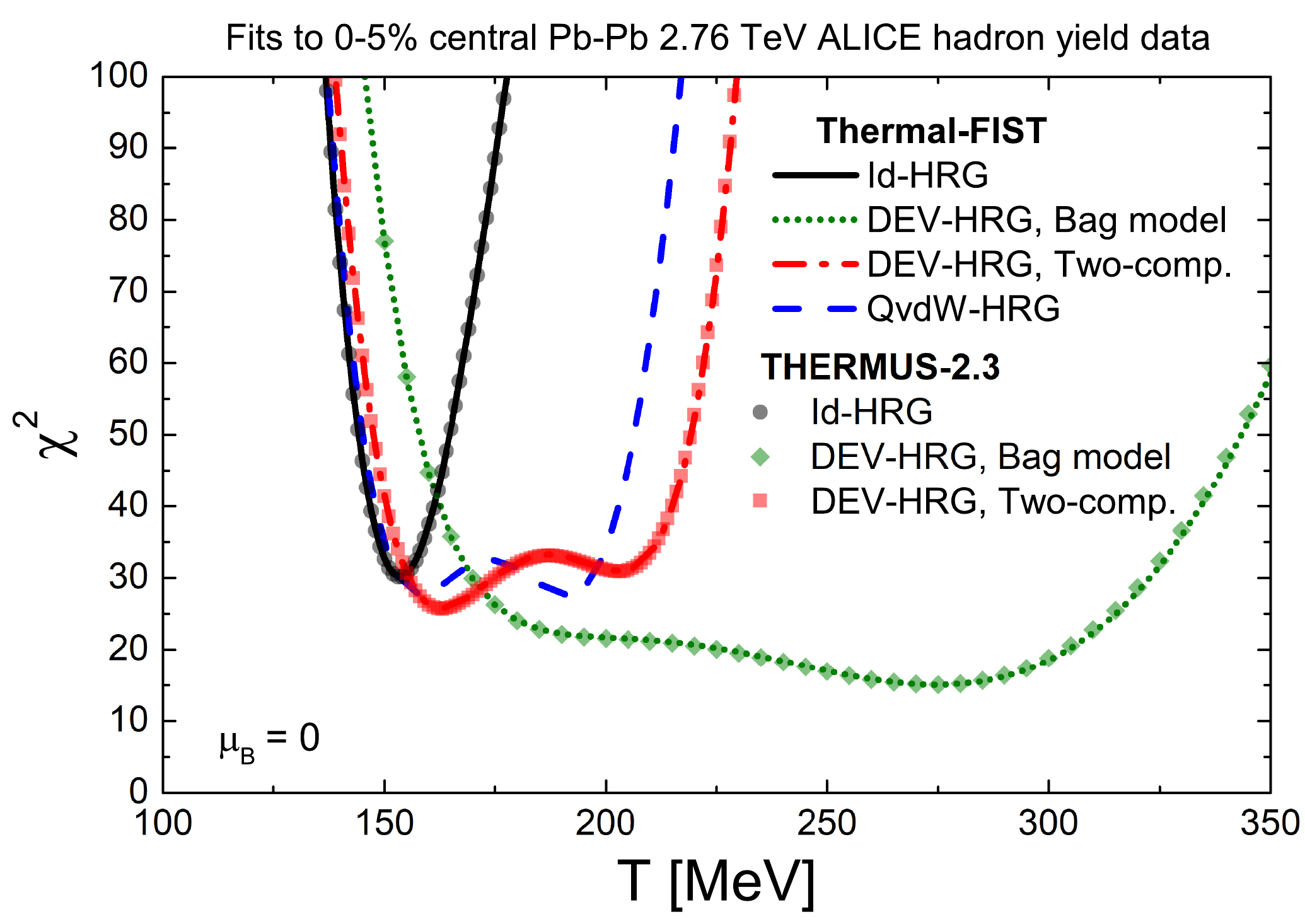}
  \caption{Temperature dependence of $\chi^2$ of the fits to hadron yields measured by the ALICE collaboration in 0-5\% central Pb-Pb collisions at $\sqrt{s_{_{NN}}} = 2.76$~TeV.
  Calculations were performed within the Id-HRG model (solid black line and circles), the DEV-HRG model with bag model parametrization of the hadron radii, with proton radius $r_p = 0.5$~fm~(dotted green line and diamonds), the DEV-HRG model with zero meson radius and $r_B = 0.3$~fm radius for baryons~(dash-dotted red line and squares), and the QvdW-HRG model~(dashed blue line).
  The lines correspond to calculations within \thermalfist, while the symbols depict \textsc{THERMUS-2.3} calculations.
  Hadron list from \textsc{THERMUS-2.3} is used in \thermalfist calculations.
  }
  \label{fig:ALICEfits}
\end{figure}

The temperature dependence of $\chi^2$ of the fit to the ALICE data, performed within the four models described above, is depicted in Fig.~\ref{fig:ALICEfits} by lines~(\thermalfist) and symbols~(\textsc{THERMUS-2.3}).
The Id-HRG model shows a single $\chi^2$ minimum at $T \sim 155$~MeV, the result reported in the literature by many groups~\cite{Petran:2013lja,Stachel:2013zma,Floris:2014pta,Becattini:2014hla}.
Calculations within the non-ideal HRG models show peculiar two-minimum structures in Fig.~\ref{fig:ALICEfits}, the mathematical origin of these structures was discussed in Refs.~\cite{Vovchenko:2015cbk,Satarov:2016peb}.
Note that physical interpretation of the high-temperature 2nd minima should be done with care, lattice QCD calculations suggest that the crossover transition to quarks and gluons may already be completed at lower temperatures~\cite{Borsanyi:2010bp,Bazavov:2011nk}.
The calculations are presented here merely in the context of code testing and comparison.
The calculations within the Id-HRG and the two EV-HRG models are quantitatively consistent between \thermalfist and \textsc{THERMUS-2.3}.
There are no \textsc{THERMUS-2.3} calculations for the QvdW-HRG model available due to a lack of its implementation in \textsc{THERMUS-2.3}.

The annotated macro, which performs the above calculations, can be found in the following location: \href{https://github.com/vlvovch/Thermal-FIST/blob/master/src/examples/cpc/cpc2-chi2-vs-T.cpp}{\texttt{src/examples/cpc/cpc2-chi2-vs-T.cpp}}

\subsubsection{Chemical non-equilibrium fits}

Previous considerations were restricted to the chemical equilibrium HRG model, i.e. for $\gamma_q = \gamma_S = 1$.
The chemical non-equilibrium scenario, $\gamma_q \neq 1$, $\gamma_S \neq 1$, for hadron production in heavy-ion collisions was advocated in Refs.~\cite{Letessier:2005qe,Petran:2013lja,Petran:2013qla}.
Fits within the chemical non-equilibrium scenario lead to significantly smaller $\chi^2 / dof$ values.
The resulting deviations from chemical equilibrium results are significant, e.g. the extracted temperature is about 15-20 MeV lower, the extracted $\gamma_q$ values are in the 1.6-1.7 range, very close to the Bose-Einstein condensation singularity for pions.
Therefore, the improvement in fit quality comes at the cost of abandoning the chemical equilibrium scenario for heavy-ion collisions. 

Here we do not discuss which physical scenario should be preferred.
Instead, we verify whether the \thermalfist package can reproduce the previously published results obtained in the chemical non-equilibrium scenario. 
To our knowledge, the chemical non-equilibrium calculations were previously restricted to the usage of \textsc{SHARE} package.

We consider the Id-HRG model fits in \thermalfist within chemical equilibrium~($\gamma_q = \gamma_S = 1$) and chemical non-equilibrium~($\gamma_q \neq 1$, $\gamma_S \neq 1$) scenarios.
The data fitted include the 4$\pi$ multiplicities reported by the NA49 collaboration~\cite{Afanasiev:2002mx,Alt:2006dk,Alt:2007aa,Alt:2008qm,Alt:2008iv,Alt:2004kq} for most central Pb-Pb collisions at $\sqrt{s_{NN}} = 7.6,\,8.8,\,12.3$, and $17.3$~GeV, and the midrapidity yields measured by the ALICE collaboration in 0-5\% most central Pb-Pb collisions at $\sqrt{s_{NN}} = 2.76$~TeV.
These data sets are consistent with the ones analyzed with \textsc{SHARE} in Refs.~\cite{Letessier:2005qe,Petran:2013lja}.

There are some technical differences between the current analysis and and previously published \textsc{SHARE} analyses.
First, the electric charge and strangeness chemical potentials $\mu_Q$ and $\mu_S$ are not fitted here, but are fixed from the conditions of the electric-to-baryon charge ratio $Q/B = 0.4$ and strangeness neutrality $S = 0$.
Second, there are differences in the hadron list employed in \thermalfist and \textsc{SHARE}, and there are also differences in the treatment of the finite resonance widths. 
Therefore, some differences between the two codes are expected, the important question is whether the qualitative features reported in Refs.~\cite{Letessier:2005qe,Petran:2013lja} can be reproduced.

\begin{figure}[t]
  \centering
  \includegraphics[width=.49\textwidth]{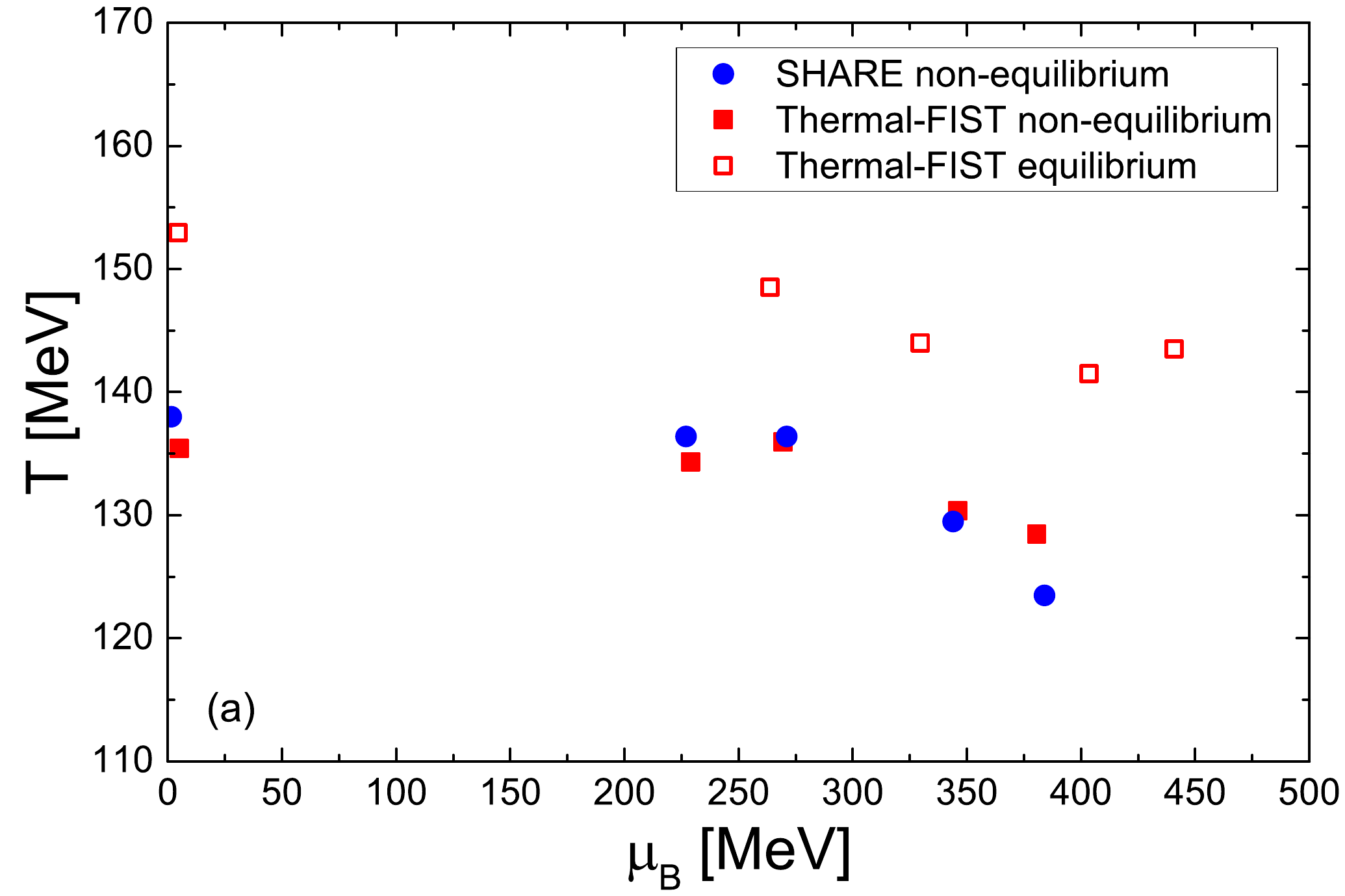}
  \includegraphics[width=.49\textwidth]{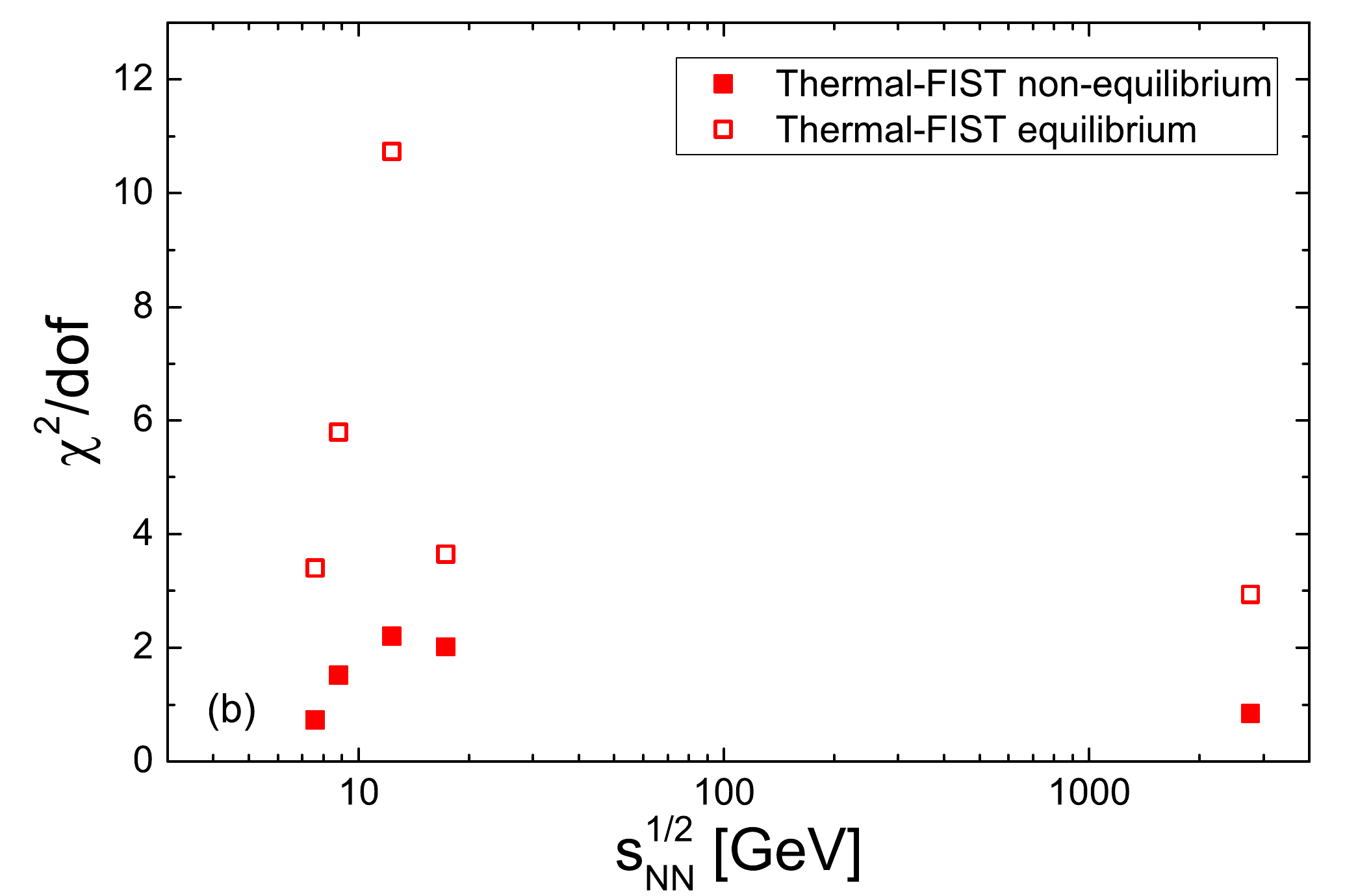}
  \caption{(a) The $T$-$\mu_B$ values extracted from the thermal fits to NA49 and ALICE data within the Id-HRG model in the chemical non-equilibrium~(full red squares) and chemical equilibrium~(open red squares) scenarios.
  Result of the chemical non-equilibrium fits performed within SHARE package from Refs.~\cite{Letessier:2005qe,Petran:2013lja} are shown by the full blue circles.
  (b) Energy dependence of the extracted $\chi^2 / dof$ values from fits  performed within the chemical non-equilibrium~(full red squares) and chemical equilibrium~(open red squares) scenarios.
  }
  \label{fig:neq}
\end{figure}

The $T$-$\mu_B$ values extracted from the thermal fits to NA49 and ALICE data within the chemical non-equilibrium  and chemical equilibrium scenarios are depicted in Fig.~\ref{fig:neq}a by full and open red squares, respectively.
The \textsc{SHARE} non-equilibrium results from Refs.~\cite{Letessier:2005qe,Petran:2013lja} are shown by the blue circles.
The \thermalfist and \textsc{SHARE} results are consistent with each other to a good precision. 
The $\gamma_q$ values extracted from fits within \thermalfist are in the range $\gamma_q = 1.6-1.7$, close to the onset of pion Bose-Einstein condensation, and consistent with the values reported in Refs.~\cite{Letessier:2005qe,Petran:2013lja}.
Figure~\ref{fig:neq}b depicts the $\chi^2 / dof$ values obtained in chemical (non-)equilibrium fits within \thermalfist.
The reduced $\chi^2$ is significantly smaller in the chemical non-equilibrium scenario, as reported in Refs.~\cite{Letessier:2005qe,Petran:2013lja}.
The presented calculation shows that \thermalfist is able to reproduce  previously published systematics of the chemical non-equilibrium scenario.

The annotated macro, which performs the above calculations, can be found in the following location: \href{https://github.com/vlvovch/Thermal-FIST/blob/master/src/examples/cpc/cpc3-chi2NEQ.cpp}{\texttt{src/examples/cpc/cpc3-chi2NEQ.cpp}}

\subsection{Analytic calculations vs Monte Carlo}

The last section here illustrates applications of the Monte Carlo TEG.
The TEG is most useful when an analytic approach is unavailable or problematic.
One such example is a simultaneous inclusion of the CE and EV effects. This procedure is described in detail in Ref.~\cite{Vovchenko:2018cnf}.
Another possibility is a study of fluctuations and correlations of various hadron yields.
Various effects, such as radial flow and kinematic cuts can be naturally included in the Monte Carlo approach whereas the whole procedure of the event-by-event analysis resembles closely the experimental situation.

In this section we consider a comparison of the analytic and Monte Carlo results for various 2nd order fluctuations and correlations of hadron yields.
Such a comparison serves as an important cross-check of consistency between analytic and Monte Carlo methods in general, and of the probabilistic decays implementation in Sec.~\ref{sec:flucdecays} in particular.

More specifically, we consider the mean-to-variance ratio of the net-kaon distribution $M_k/\sigma_2^k$, 
the ratio $\sigma_{11}^{Qk} / \sigma_2^k$ of the correlator between the total electric charge and final net kaon number over the variance of the final net kaon number,
the ratio $\sigma_{11}^{pQ} / \sigma_2^p$ of the correlator between the final proton number and the total electric charge  over the variance of the final net proton number, and
the ratio $\sigma_{11}^{pk} / \sigma_2^k$ of the correlator between the final proton number and the final net kaon number over the variance of the final net kaon number, in the framework of the Id-HRG model.
The beam energy dependence of $M_k/\sigma_2^K$ measured in Au-Au collisions was recently published by the STAR collaboration~\cite{Adamczyk:2017wsl}, while the experimental analysis of the correlation observables is ongoing.

\begin{figure}[t]
  \centering
  \includegraphics[width=\textwidth]{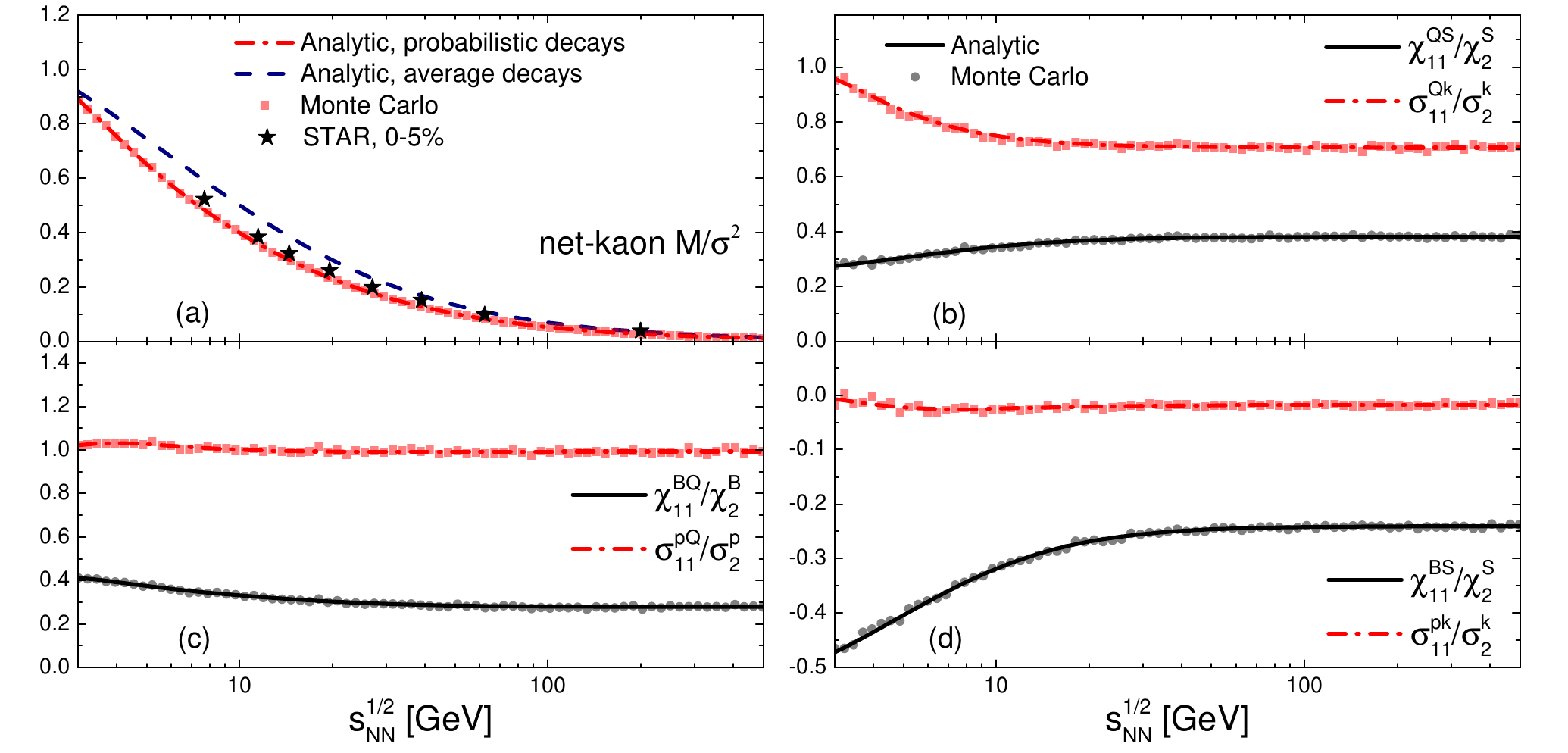}
  \caption{
 Collision energy dependence of  (a) Net-kaon mean-to-variance ratio, (b) susceptibility ratios $\chi_{11}^{QS} / \chi_2^S$ and $\sigma_{11}^{Qk} / \sigma_2^K$, (c) $\chi_{11}^{BQ} / \chi_2^B$ and $\sigma_{11}^{pQ} / \sigma_2^p$,
  (d) $\chi_{11}^{BS} / \chi_2^S$ and $\sigma_{11}^{pk} / \sigma_2^k$ calculated within the Id-HRG model using analytical (lines) and Monte Carlo (symbols) methods.
  The stars in panel (a) depict the STAR data for net-kaon $M/\sigma^2$ for 0-5\% central Au-Au collisions~\cite{Adamczyk:2017wsl}.
  }
  \label{fig:BESsusc}
\end{figure}

The quantities $\sigma_2^k$, $\sigma_2^p$, $\sigma_{11}^{Qk}$, $\sigma_{11}^{pQ}$, and $\sigma_{11}^{pk}$ can be expressed as linear combinations over the correlators~\eqref{eq:Nijcorr} of the number of final state hadrons, which are calculated within the \thermalfist.
The \thermalfist calculations here are performed along the chemical freeze-out curve of Ref.~\cite{Cleymans:2005xv}, the resulting beam energy dependences are shown in Fig.~\ref{fig:BESsusc} by red lines.
The red lines in Fig.~\ref{fig:BESsusc} correspond to analytic treatment of fluctuations, including the effects of probabilistic decays.

The calculation results for the same observables using the Monte Carlo TEG are shown by the red symbols. These are consistent with the analytic results, thus verifying the accuracy of the probabilistic decay treatment implemented in \thermalfist. In contrast, the simplified ``average'' decay treatment procedure~\cite{Nahrgang:2014fza} leads to a markedly different result for $M_k/\sigma_2^K$~[dashed black curve in Fig.~\ref{fig:BESsusc}(a)] in comparison to probabilistic decays and to Monte Carlo result.
Full probabilistic treatment of resonance decays thus appears to be important for interpretation of the corresponding data within the HRG approach.

The quantities $\sigma_{11}^{Qk} / \sigma_2^K$,  $\sigma_{11}^{pQ} / \sigma_2^p$, and $\sigma_{11}^{pk} / \sigma_2^k$ are considered as possible proxies for the corresponding ratios of susceptiblities of conserved charges, $\chi_{11}^{QS} / \chi_2^S$, $\chi_{11}^{BQ} / \chi_2^B$, and $\chi_{11}^{BS} / \chi_2^S$. The \thermalfist calculations for the latter are shown in Fig.~\ref{fig:BESsusc} by the black lines (analytical) and symbols (Monte Carlo), and these differ substantially from the measurable correlator ratios. These significant differences should be considered when interpreting the experimental data in the context of the equation of state of QCD matter.

The annotated macro, which performs the above calculations, can be found in the following location: \href{https://github.com/vlvovch/Thermal-FIST/blob/master/src/examples/cpc/cpc4-mcHRG.cpp}{\texttt{src/examples/cpc/cpc4-mcHRG.cpp}}

\section{Concluding remarks}
\label{sec:summary}

The \thermalfist package provides the ability to perform an analysis of both, the hadronic part of the QCD equation of state and of the statistical description of hadron production in heavy-ion collisions.

The powerful graphical user interface in the package will be useful for intepreting the future hadron yield data which will come in the ongoing and future heavy-ion experiments.

Generalization and development of different variants of the HRG model is also presently an active research topic.
In that regard, the flexible modular structure of \thermalfist
is particulary suitable for continuous improvement of the package and implementation of various new features to test different new ideas.

\section*{Acknowledgments}

We would like to thank P. Alba, D. Anchishkin, V. Begun, E. Bratkovskaya, B.~D\"onigus, M. Gorenstein, I. Kisel, M. Lorenz, D. Oliinychenko, and L. Satarov for fruitful discussions and physics suggestions.
We also acknowledge useful feedback from F.~Flor, R.~Hensch, A.~Motornenko, R.~Poberezhnyuk, P.~Parotto, M.~Puccio, and J.~Stumm.
H.St. acknowledges the support through the Judah M. Eisenberg Laureatus Chair at Goethe University, and the Walter Greiner Gesellschaft, Frankfurt.

\section*{References}

\begingroup
\renewcommand{\section}[2]{}%
\bibliographystyle{elsarticle-num}
\bibliography{thermal-fist.bib}
\endgroup

\end{document}